\documentclass[a4paper,11pt]{article}
\usepackage{jheppub} % for details on the use of the package, please
\usepackage{amsthm} %for dealing with theorem environments,

\usepackage{stmaryrd}

\newcommand{\D}{\Delta}
\newcommand{\hD}{\Delta}
\newcommand{\hM}{M}

\newcommand{\hB}{B}
\newcommand{\hb}{b}

\newcommand{\hL}{L}

\newcommand{\hW}{W}
\newcommand{\iw}{\frac{1}{w}}
\newcommand{\iz}{\frac{1}{z}}
\newcommand{\iwz}{\frac{1}{wz}}
\newcommand{\nn}{\nonumber}

\newtheorem*{thm}{Theorem}
\newtheorem*{thm_A}{Theorem A}
\newtheorem*{thm_B}{Theorem B}
\newtheorem*{lem}{Lemma}
\newtheorem*{lem_A}{Lemma A}
\newtheorem*{lem_B}{Lemma B}
\newtheorem*{dfn}{Definition}

\newtheorem*{cor_1}{Corollary 1}
\newtheorem*{cor_2}{Corollary 2}

\title{General solution of  cyclic Leibniz rule}

\author[a]{Daisuke Kadoh}
\author[b]{and Naoya Ukita}

\affiliation[a]{KEK Theory Center, High Energy Accelerator Research Organization (KEK),\\
            \ Tsukuba, Ibaraki 305-0801, Japan}
\affiliation[b]{Center for Computational Sciences, University of Tsukuba,\\ 
            \ Tsukuba, Ibaraki 305-8577, Japan}
\emailAdd{kadoh@post.kek.jp}
\emailAdd{ukita@ccs.tsukuba.ac.jp}

\preprint{
{\flushright
 \vspace{-8.5mm}
KEK-TH-1756, UTCCS-P-75
\\
}}

\abstract
{
We study the general solution of the cyclic Leibniz rule (CLR) which was recently proposed as a new approach to the lattice supersymmetry. 
Introducing some mathematical preliminaries related to the cyclic symmetry, we find the general solution of the 2-body CLR for the naive symmetric difference operator. 
The main theorems of this paper state that  the general solution can be uniquely expressed as (A)  a linear combination of the two fundamental solutions with cyclic invariant coefficients, and (B) a linear combination of the minimal solutions with complex coefficients. 
Moreover, an extension to the general difference operators is also discussed.
}


\begin{document}
\maketitle
%\flushbottom

%
% Introduction
%
\section{Introduction}
\label{sec:intro}
%
% Motivation of SUSY
Supersymmetry (SUSY) has become an important focus in elementary particle physics for several reasons. 
It was originally introduced as a generalization of the symmetry of the S-matrix consistent with relativistic quantum field theory \cite{Haag},
and has been studied as a strong candidate for physics beyond the standard model and an important ingredient of superstring \cite{Green, Polchinski}. 
Furthermore, for supersymmetric models, some exact results can be obtained  and give us great insights into quantum field theory \cite{Seiberg, Seiberg-Witten, Nekrasov, Pestun}.   
However, we do not fully understand the dynamics of supersymmetric field theory in general.% which still remains as a problem.

Lattice theory is a powerful tool to reveal the dynamics of field theory. 
Indeed lattice gauge theory achieves great success in lattice QCD,
 but the success is never confined to QCD and will be also realized in supersymmetry. 
It is, however, difficult to put supersymmetry on the lattice due to the violation of the Leibniz rule \cite{Kato-1}. 
The realization of the SUSY continuum limit using a fine-tuning of the SUSY breaking operators is also difficult in practice, so that
the lattice supersymmetry has been a difficult problem.

There are several attempts to overcome the problem. Most of them realize a subset of the SUSY transformations on the lattice such that the full supersymmetry is restored in the continuum limit without any fine-tunings. 
In \cite{Elitzur, Cecotti, Sakai, Kikukawa}, ${\mathcal N}=2$ Wess-Zumino model in $D=2$ is constructed on the basis of the Nicolai mapping.  Furthermore, supersymmetric Yang-Mills lattice theories are given by the orbifolding technique \cite{Kaplan, Cohen} and the methods based on topological field theory \cite{Sugino, Catterall}. 
Those attempts look different to each other, but in fact they are the same in the sense that a few nilpotent supercharges $Q$ are realized on the  lattice, the corresponding actions are given by the $Q$-exact forms,       
and the $Q$-invariance is manifest from the nilpotency. 
The lattice models in $D\leq2$ achieve the correct continuum limit without any fine-tunings. But, in $D=4$, there have been no satisfactory solutions with  an exception of ${\mathcal N}=1$ supersymmetric Yang-Mills theory so far. 
Therefore, we need a new method that is conceptually different from the previous attempts. 

Recently, an entirely different approach was proposed by Kato, Sakamoto and So \cite{Kato-2}, who found that the cyclic Leibniz rule (CLR) for difference operators, instead of the usual Leibniz rule, is significant.  Using the CLR, they  defined supersymmetric quantum mechanics on the lattice, which preserves a subset of the SUSY algebra, $Q^2=0$. 
The exact charge is the same as that of the previous works, but
the CLR has better properties:  the Witten index can be analytically calculated in the same manner as in the continuum theory. The model with the CLR does not have the surface terms which exist in the old models and prevent us from realizing the larger SUSY transformations.  
In addition, the kinetic terms and the potential terms are independently invariant.
In \cite{Kato-2}, some solutions of the CLR and the general form of the product of the solutions  have been already given, 
\footnote{
In the notation of (\ref{CLR_w_rep}), the general form of the product of $M$ and $\Delta$ is given in terms of a two variable holomorphic function $f(w,z)$.
}
nevertheless further detailed studies are required for an understanding of the structure of the solutions.

In this paper, we construct the general solution of the 2-body CLR for the naive symmetric difference operator and reveal its structure.
We review the CLR in section 2. Some important results and mathematical preliminaries are given in section 3, after which we show our main theorems in section 4. 
Section 5 is  devoted to a summary. In appendix A, the polynomial solution of degree six is explicitly given.  
The general polynomial solution of any degree is shown in appendix B.  
An extension to  the general difference operators is discussed in appendix C.

%%%%%%%%%%%%%%%%%%%%%%%%%%%%%%%%%%%%%%%%%%%%%%%%%%%%%%%%%%%%%%%%%%%%
%                                                                                                   %
%                                                                                                   %
%                                                                                                   %
%                     Cyclic Leibniz rule                                                      % 
%                                                                                                   %
%                                                                                                   % 
%                                                                                                   %
%%%%%%%%%%%%%%%%%%%%%%%%%%%%%%%%%%%%%%%%%%%%%%%%%%%%%%%%%%%%%%%%%%%%

\section{Cyclic Leibniz rule}
We review the cyclic Leibniz rule 
 for supersymmetric quantum mechanics, proposed in \cite{Kato-2}.

In the continuum theory \cite{Witten}, 
supersymmetric quantum mechanics consists of a supermultiplet $(\phi(t),\psi(t),F(t))$, where 
$\phi(t)$ is a real scalar, $F(t)$ is a real auxiliary field, $\psi(t)$ is a complex fermion and $t$ is Euclidean time.
The action is given by
\begin{eqnarray}
 S_{\rm SQM}
 =
 \int dt \left\{
   \frac{1}{2}\left(\frac{d\phi}{dt}\right)^2+\frac{1}{2}F^2+iFW(\phi)
   +i\bar{\psi}\left(\frac{d}{dt}+\frac{\partial W(\phi)}{\partial\phi}\right)\psi
 \right\},
 \label{SQM_con}
\end{eqnarray}
where $W(\phi)$ is an arbitrary function of $\phi$.  

The action (\ref{SQM_con}) is invariant under the following SUSY transformations,
\begin{eqnarray}
\label{SUSY_con_s}
 \delta\phi &=& \epsilon\bar\psi-\bar\epsilon\psi,\\
 \delta\psi &=& \epsilon\left(i\frac{d\phi}{dt}+F\right),\\
 \delta\bar\psi &=&\bar\epsilon\left(-i\frac{d\phi}{dt}+F\right),\\ 
\label{SUSY_con_e}
 \delta F &=& -i\epsilon\frac{d\bar\psi}{dt}-i\bar\epsilon\frac{d\psi}{dt},
\end{eqnarray}
where  $\epsilon$, $\bar\epsilon$ are two Grassmann  parameters. 
Supercharges $Q$ and  $\bar Q$, which are defined by $\delta = \epsilon Q + \bar\epsilon\bar Q$, satisfy the well-known SUSY algebra, 
\begin{eqnarray}
 \{Q, Q\} =  \{\bar Q, \bar Q\} =0,\ \ 
 \{\bar Q, Q\} = -2i\frac{d}{dt}.
 \label{SUSY_alg_con}
\end{eqnarray}
The Leibniz rule for the differential operator $\frac{d}{dt}$ in (\ref{SQM_con})--(\ref{SUSY_alg_con}) leads to the SUSY invariance of the action. 

In order to construct a lattice model, let us discretize Euclidean time   
as a one dimensional lattice. We set the lattice spacing $a=1$ for simplicity and label the lattice sites as $n \in {\mathbb Z}$. Then the  supermultiplet $(\phi_n, \psi_n, F_n)$ lives on the sites.  
Hereafter, let us consider a $\phi^4$ theory given by  
$W(\phi)=m\phi+\frac{g}{2}\phi^2$ with a mass $m$ and a coupling constant  $g$.

In \cite{Kato-2}, 
the following lattice action $S_{\rm SQM}^{\rm lat}=S_0 + S_m +S_{\rm int}$ is introduced,
\begin{eqnarray}
 S_0 &=& \sum_{n} \left(\frac{1}{2}(\D\phi)^2_n+i\bar\psi_n(\D\psi)_n+\frac{1}{2}F^2_n\right),\label{Skin}\\
 S_m &=& i\sum_{m,n}G_{mn}\left(F_m\phi_n+\bar\psi_m\psi_n\right),\label{Smass}\\
 S_{\rm int} &=&
 i\frac{g}{2}\sum_{l,m,n}M_{lmn}(F_l\phi_m\phi_n+2\phi_n\bar\psi_m\psi_l),
 \label{SintQ}
\end{eqnarray}
where 
$S_0$ is the kinetic term with a difference operator $\D$ 
% instead of the differential operator $\frac{d}{dt}$, 
acting on a field $\phi$ as
\begin{eqnarray} 
 (\D\phi)_n \equiv \sum_{m}\D_{nm}\phi_m,
 \label{difference-operator_def}
\end{eqnarray}
and $S_m$ is the mass term with the Wilson term given by 
\begin{eqnarray}
 G_{mn} = m\delta_{m,n} + r\frac{2\delta_{m,n}-\delta_{m,n-1}-\delta_{m,n+1}}{2},
\end{eqnarray}
where $r$ is the Wilson parameter which is non-zero if $\D$ has doublers.   
$S_{\rm int}$ is the cubic interaction term with the coupling  constant $g$ and coefficients $M_{lmn}$ that give the interaction on the lattice.\footnote{
Instead of (\ref{SintQ}), we may use  
\begin{eqnarray}
 S_{\rm int} = i\frac{g}{2}\sum_{l,m,n}
(M_{lmn}F_l\phi_m\phi_n +2N_{lmn}\phi_l\bar\psi_m\psi_n),
\end{eqnarray}
 where $M_{lmn}=M_{lnm}$.
 The lattice $Q$ transformations (\ref{SUSY_lat_s})-(\ref{SUSY_lat_e})  yield $M_{lmn}=N_{nml}$. We used this in (\ref{SintQ}) on ahead for simplicity.
}

 To achieve the SUSY invariance, we employ the   
 symmetric difference operators $\D_{mn}$ satisfying \footnote{
 The difference operators can be formally defined by 
 $\sum_{m}\D_{nm}=0$ which means that $\D$ acting on any constant vanishes. 
 The translational invariance, $\D_{nm}=\D_{0,m-n}$, tells us that
 the symmetric difference operators satisfy this condition.}
\begin{eqnarray} 
 \D_{mn} = -\D_{nm},
 \label{difference-operator_sym}
\end{eqnarray}
and  assume that $M_{lmn}$ 
   satisfy the 2-body CLR,\footnote{
   In \cite{Kato-2}, 
the Leibniz rule (LR) on the lattice is given by
\begin{eqnarray}
\sum_{k}(M_{kmn}\D_{kl}+M_{lkn}\D_{km}+M_{lmk}\D_{kn})=0.
\end{eqnarray}
Note that how to contract the indices of $M_{lmn}$ is different from that in the CLR (\ref{CLR_x_rep}). 
%Using the field product, $\{\phi,\psi\}_l\equiv\sum_{mn}M_{lmn}\phi_m\psi_n$, and the inner product, $(\phi,\psi)\equiv\sum_{n}\phi_n\psi_n$, in \cite{Kato-2}, the CLR can be also written as 
%\begin{eqnarray}
%(\hD\phi,\{\psi,\chi\})+(\hD\psi,\{\chi,\phi\})+(\hD\chi,\{\phi,\psi\})=0,
%\end{eqnarray}
%while  
%the LR is
%\begin{eqnarray}
%(\hD\phi,\{\psi,\chi\})+(\phi,\{\hD\psi,\chi\})+(\phi,\{\psi,\hD\chi\})=0.
%\end{eqnarray}
Although the CLR and the LR are different to each other at any finite lattice spacing, but both reproduce the same continuum LR in the continuum limit.}
\begin{eqnarray}
 \sum_{k}(M_{kmn}\D_{kl}+M_{knl}\D_{km}+M_{klm}\D_{kn})=0,
 \label{CLR_x_rep}
\end{eqnarray}
where $M_{lmn}$ are symmetric for the last two indices,
\begin{eqnarray}
 M_{lmn}=M_{lnm}.
 \label{M_cond}
\end{eqnarray}
In the continuum limit, we also assume that $\D\rightarrow\frac{d}{dt}$ and $M_{lmn}\rightarrow\delta_{lm}\delta_{ln}$, and the lattice model (\ref{Skin})--(\ref{SintQ}) reproduces the correct continuum theory (\ref{SQM_con}). 

Furthermore, we assume the translational invariance, 
\begin{eqnarray}
&&\D_{mn}=\D_{0,n-m}, \label{D_inv_trans}\\
&&M_{lmn}=M_{0,m-l,n-l}, \label{M_inv_trans}
\end{eqnarray}
 and 
the exponential locality conditions, 
\begin{eqnarray}
 &&|\D_{mn}| \leq C \exp(-K|m-n|),\label{local_D}
 \\
 &&|M_{lmn}| \leq C \exp(-K|m-l|-K|n-l|),
 \label{local_M}
\end{eqnarray}
where $C, K$ are positive constants.

The lattice SUSY transformations for the multiplet $(\phi_n, \psi_n, F_n)$ are defined by 
\begin{eqnarray}
\label{SUSY_lat_s}
 \delta\phi_n &=& \epsilon\bar\psi_n-\bar\epsilon\psi_n,\\
 \delta\psi_n &=& \epsilon(i(\D\phi)_n+F_n),\\
 \delta\bar\psi_m &=& \bar\epsilon(-i(\D\phi)_n+F_n),\\ 
\label{SUSY_lat_e}
 \delta F_n &=& -i\epsilon(\D\bar\psi)_n-i\bar\epsilon(\D\psi)_n,
\end{eqnarray}
and the corresponding supercharges $Q$ and $\bar Q$ satisfy a lattice counterpart of the SUSY algebra (\ref{SUSY_alg_con}),
\begin{eqnarray}
 \{Q, Q\} =  \{\bar Q, \bar Q\} =0,\ \ 
 \{\bar Q, Q\} = -2i\D.
 \label{SUSY_alg_lat}
\end{eqnarray}
As discussed below, the lattice action  (\ref{Skin})-(\ref{SintQ}) is invariant under the $Q$-transformation.

In the case of the symmetric difference operators, 
the free action $S_0$ (\ref{Skin}) and $S_m$ (\ref{Smass}) are invariant under both $Q$ and $\bar Q$ transformations (\ref{SUSY_lat_s})--(\ref{SUSY_lat_e}). 
Meanwhile the interaction term $S_{\rm int}$ (\ref{SintQ}) is not invariant under these two transformations. The $Q$ transformation of $S_{\rm int}$ is proportional to the CLR,
\begin{eqnarray}
 Q S_{\rm int} &=& 
 \frac{g}{2}\sum_{l,m,n}\sum_{k}(M_{kmn}\D_{kl}+M_{knl}\D_{km}+M_{klm}\D_{kn})
                          \psi_l\phi_m\phi_n,
\end{eqnarray}
 and vanishes.
However, the $\bar Q$ transformation of $S_{\rm int}$ is not so, in general.
%Hence, $S_{\rm int}$ possesses the $Q$-invariance.  

To realize the $\bar Q$-invariance, the interaction term should be changed to the form, 
\begin{eqnarray}
 S_{\rm int}^{\prime} = i\frac{g}{2}\sum_{l,m,n}M_{lmn}(F_l\phi_m\phi_n+2\phi_n\bar\psi_l\psi_m).
 \label{SintbQ}
\end{eqnarray}
Indeed, $\bar Q S^{\prime}_{\rm int}$ is proportional to the CLR.  
If $M_{lmn}$ are totally symmetric, then 
$S_{\rm int}^{\prime}$ (\ref{SintbQ}) is identical to  $S_{\rm int}$ (\ref{SintQ}), so that both $Q$ and $\bar Q$ are realized on the lattice.
But the totally symmetric and local $M_{lmn}$ that satisfy the CLR lead to the usual Leibniz rule, and such $M_{lmn}$ do not exist from the no-go theorem \cite{Kato-1}.    
Therefore,  either $Q$ or $\bar Q$ becomes the exact charge for the interacting theory.

In this sense, the number of supercharges realized on the lattice is the same as that of the previous attempts using the Nicolai mapping. 
However, this construction has advantages over the other lattice models.
The Witten index can be analytically estimated in the same manner as in the continuum theory. 
In the ${\mathcal N}=(2,2)$ lattice Wess-Zumino model with the Wilson fermion \cite{Elitzur, Cecotti, Sakai}, holomorphy is broken by the surface terms. But the terms are forbidden by the CLR, so that one can expect that holomorphy is kept on the lattice \cite{Kato-3}.     
In addition, the kinetic terms and the potential terms are independently invariant  under the partial SUSY transformation.

Before we end this section,  we show a simple solution of the CLR with   translational invariance and locality, derived in \cite{Kato-2}.
The solution is given by
\begin{eqnarray}
 \D_{mn} &=& \frac{1}{2}(\delta_{m.n-1}-\delta_{m,n+1}),\label{sol_D}\\
 M_{lmn} &=& \frac{1}{6}(2\delta_{l,m-1}\delta_{l,n-1}
                       +\delta_{l,m+1}\delta_{l,n-1}
                       +\delta_{l,m-1}\delta_{l,n+1}
                      +2\delta_{l,m+1}\delta_{l,n+1}).\label{sol_M}
\end{eqnarray}
In the naive continuum limit, $\D\rightarrow \frac{d}{dt}$, $M_{lmn}\rightarrow \delta_{lm}\delta_{ln}$. 
Note that $\D_{mn}$ (\ref{sol_D}) and $M_{lmn}$ (\ref{sol_M}) have compact support, which  are referred  to as ultra-locality, and of course satisfy  exponential locality (\ref{local_D}), (\ref{local_M}).

%
% Preparation for the general solution of CLR
%
\vspace{5mm}
\section{Mathematical preliminaries and some results}
Before considering the general solution of the CLR, we present some mathematical preliminaries and results in  this section.
Starting from an introduction to a new locality condition that respects the cyclic symmetry in section 3.1, after which  
we give the CLR in the momentum space (or on "cyclic invariant annulus") in section 3.2.  Spaces of the solutions and cyclic invariant functions  are defined in section 3.3 and a new definition of the degree of a polynomial solution is also introduced. 
In section 3.4, expansions of the cyclic invariant functions are given.   
We derive  a recurrence relation related to the CLR and, by solving it,  obtain the several solutions of lower degree in section 3.5. 
We define mappings referred to as delta operators in section 3.6, which are important in the proof of the theorem.
 In the last section 3.7, we give particular solutions  of the CLR, which are referred to as minimal solutions.

%%%%%%%%%%%%%%%%%%%%%%%%%%%%%%%%%%%%%%%%%%%%%%%%%%%%%%%%%%%%%%%%%%%%
%                                                                                                   %
%                                                                                                   %
%                                                                                                   %
%                     Locality and cyclic symmetry                                        % 
%                                                                                                   %
%                                                                                                   % 
%                                                                                                   %
%%%%%%%%%%%%%%%%%%%%%%%%%%%%%%%%%%%%%%%%%%%%%%%%%%%%%%%%%%%%%%%%%%%%
\vspace{5mm}
\subsection{Locality and cyclic symmetry}
%
%   Introduction of this subsection
%
The exponential  locality condition (\ref{local_M}) and the cyclic symmetry do not go well together. 
In this section, let us define a locality condition compatible with the cyclic symmetry for a mathematically beautiful construction of the theory of the CLR. 

%
%  Cyclic transformation
%
Let $O_{lmn}$ be any quantity with three indices, $l,m,n \in {\mathbb Z}$. 
Then, the cyclic transformation of $O_{lmn}$ is defined by the permutation of the three indices,
\begin{eqnarray}
 {\cal O}_{lmn} \rightarrow \ {\cal O}_{mnl}.
 \label{cyclic_trans}
\end{eqnarray}
Note that applying this transformation three times produces the identity mapping. 
If we choose the left-hand side of  the CLR (\ref{CLR_x_rep}) as $O_{lmn}$,
\begin{eqnarray}
 {\cal O}_{lmn}= \sum_{k} ( M_{kmn}\D_{kl}+M_{knl}\D_{km}+M_{klm}\D_{kn}),
\end{eqnarray}
as its name suggests, 
the CLR is invariant under the cyclic transformation (\ref{cyclic_trans}).

%
%  Cyclic transformation of S_kl
%
Now let us consider the cyclic transformation of a solution  of the CLR, which is denoted by $M_{lmn}$. 
From the translational invariance (\ref{M_inv_trans}),  the symmetric matrix $S_{kl}=M_{0kl}$ gives  the independent components of $M_{lmn}$, 
and the cyclic transformation reads
\begin{eqnarray}
   S_{kl} \rightarrow S_{l-k,-k}.
   \label{S_transf}
\end{eqnarray}
The locality condition (\ref{local_M}) for $S_{kl}$ becomes 
\begin{eqnarray}
 |S_{kl}| \leq C \exp \{ -K ( |k| + |l|) \},
 \label{S_local}
\end{eqnarray}
which is not invariant under the cyclic transformation (\ref{S_transf}). 
In later sections, we will find that the cyclic non-invariant locality  condition (\ref{S_local}) is not suited for the construction of our theorems.

%
%   Equivalence relation and equivalence class   
%
In order to define a locality condition compatible with the cyclic symmetry,
let us consider an absolute value,
\begin{eqnarray}
 \vert\vert \, S_{kl} \, \vert\vert = {\rm max} \big\{ \, |S_{k,l}|,\,  |S_{l-k,-k}|, \,  | S_{-l,k-l}| \,  \big\},
 \label{equiv_class_abs}
\end{eqnarray}
which is invariant under the cyclic transformation. 
%let us consider an equivalence relation in which the set of indices $(k,l)$ are equivalent if they are changed to each other under the cyclic transformation (\ref{S_transf}),
%\begin{eqnarray}
% (k,l)  \sim (l-k,-k),
% \qquad {\rm for} \ \  k,l \in {\mathbb Z}. 
%\end{eqnarray}
%Then, the corresponding equivalence class 
%is defined by
%\begin{eqnarray}
%  [ \, (k,l ) \, ] = 
% \{ \ (i,j) \ |  \ 
% (i,j) \sim (k,l), \ {\rm for\ } i,j \in {\mathbb Z \ }
% \},  \qquad {\rm for\ \ }  k,l \in{\mathbb Z},
% \label{equiv_class}
%\end{eqnarray}
%
%
%
%with an associated absolute value, 
%\begin{eqnarray}
%  \vert\vert {\cal O}_{kl} \vert\vert = 
% \{ \ { \rm max} \{ | {\cal O}_{ij}| \} \ | \ 
% (i,j) \in [(k,l)] , \ {\rm for\ } i,j\in{\mathbb Z \ }
% \}, \qquad {\rm for\ \ }  k,l \in{\mathbb Z}, 
%\end{eqnarray}
%where ${\cal O}_{kl}$ is any quantity that transforms like (\ref{S_transf}).
%This absolute value is invariant under the cyclic transformation. 
%
%
Using (\ref{equiv_class_abs}), 
we give  a new locality condition, 
\begin{eqnarray}
 \vert\vert S_{ k, -l} \vert\vert  \le C\exp \{ -K(k+l) \}, \qquad {\rm for} \ \  k,l \in {\mathbb N}, 
 \label{cyc_local}
\end{eqnarray}
for some positive constants $C, K$.\footnote{
In this paper, ${\mathbb N}$ denotes non-negative integers, $0,1,2,3,\cdots$.} 
The cyclically equivalent elements,
\begin{eqnarray}
 \left\{ S_{k,-l}, S_{-k-l,-k},  S_{l,k+l}\right\},
 \label{equiv_class_s}
\end{eqnarray}
are equally treated and  
all of the elements are thus exponentially suppressed in the cyclic invariant way at large distances. 
This locality condition (\ref{cyc_local}) will be referred to as cyclic invariant exponential locality condition, or simply $cyclic \ locality$ in the following.

Locality is one of the fundamental requirements of the lattice theory, so  one might think that this reformulation of the locality condition is not theoretically well-supported. However the locality condition (\ref{cyc_local}) also guarantees that the interactions decay exponentially at large distances. In addition, the following inclusion relation, % between the cyclic locality  (\ref{cyc_local}) and the simple exponentially locality (\ref{S_local}),
\begin{align}  
& \nonumber \\[-12pt]
& \{ S_{kl} {\rm \ satisfying  \ the \  locality \ condition \ } (\ref{S_local}) {\rm \ with\ } K \}  
\nonumber \\[8pt]
& \subset \{ S_{kl}  {\rm \ satisfying \ the \ cyclic \  locality \ condition \ } (\ref{cyc_local}) {\rm \ with\ } K \}  
\\[8pt]
& \subset \{ S_{kl}  {\rm \ satisfying \ the \  locality \ condition \ } (\ref{S_local}) {\rm \ with\ } K/2 \} 
\nonumber, \\[-12pt] \nonumber
\end{align}
tells us that
the cyclic locality with a rate of decay $K$ is a locality condition between usual ones with the rates $K/2$ and $K$. 
Therefore, the cyclic locality maintains the  principle of locality.

The cyclic locality is introduced for multiple indices, while we keep the exponential locality condition (\ref{local_D}) for the  difference operators,
\begin{eqnarray}
 |d_{l}| \leq C \exp(-K|l|),\label{local_d}
\end{eqnarray}
where $d_{l}\- = \D_{0l}$ and   some constants $C, K >0$.
With the conditions (\ref{cyc_local}) and  (\ref{local_d}), we will construct the general solution of the CLR  (\ref{CLR_x_rep}) in later sections.

%%%%%%%%%%%%%%%%%%%%%%%%%%%%%%%%%%%%%%%%%%%%%%%%%%%%%%%%%%%%%%%%%%%%
%                                                                                                   %
%                                                                                                   %
%                                                                                                   %
%                     CLR on cyclic invariant annulus                                     % 
%                                                                                                   %
%                                                                                                   % 
%                                                                                                   %
%%%%%%%%%%%%%%%%%%%%%%%%%%%%%%%%%%%%%%%%%%%%%%%%%%%%%%%%%%%%%%%%%%%%
\vspace{5mm}
\subsection{CLR on cyclic invariant annulus}
\label{sec:cyclic_annulus}
In this section, we give the CLR in the momentum space. In particular, 
the CLR can be naturally defined on an annulus associated with the cyclic symmetry.

The first indices of $\D_{mn}$ and $M_{lmn}$ can be set to zero by the translational invariance. 
By performing the Fourier transformations of $d_{l}\equiv\D_{0l}$ and $S_{kl}\equiv M_{0kl}$, we obtain their {\it w}-representations \cite{Kato-1},
\begin{eqnarray}
 &&\hD(w)   \equiv \sum_{l\in{\mathbb Z}}  d_{l} w^l,\label{hD}\\
 &&\hM(w,z) \equiv \sum_{k,l\in{\mathbb Z}}  S_{kl} w^k z^l,\label{hM}
\end{eqnarray}
where $w, z$ are the coordinates of the unit circle $S^1$, $w=\exp(ip), z=\exp(iq)$ with the momenta $p, q \in{\mathbb R}$.

The CLR is  given by
\begin{eqnarray}
 \hM(w,z)\hD\left(\iwz\right)+\hM\left(z,\iwz\right)\hD(w)+\hM\left(\iwz,w\right)\hD(z)=0,
 \label{CLR_w_rep}
\end{eqnarray}
in the momentum space.%
\footnote{
The Leibniz rule is expressed as $\hM(w,z)\hD\left(\iwz\right)+\hM(w,z)\hD(w)+\hM(w,z)\hD(z)=0$. The no-go theorem \cite{Kato-1} tells us that the non-trivial solution with $\hM(w,z)\neq0$ is $\hD(w)\sim\log w$ which is not holomorphic on the domain ${\mathcal D_K}$, so that $\D_{mn}$ is non-local.}
\footnote{
It is valuable to comment on the zero points of $\hD(w)$ for $|w|=1$.
(\ref{cond_hD_1})  suggests that $\hD(w)$ has an even number of zeros, $w=\pm1$ at least and  the pairs $w=\rho_i, \rho_i^{-1} \ (\rho_i\neq\pm1, i=1,2,\cdots)$ 
if the others exist.
The simplest case is the naive symmetric difference operator, which has only two zero points, $w=\pm1$. 
}
 For the SUSY invariance we focus on the specific case,
\begin{eqnarray}
 &&\hD(w) = -\hD\left(w^{-1}\right),\label{cond_hD_1}\\
 &&\hM(w,z) = \hM(z,w).\label{cond_hM_1}
\end{eqnarray}
The first condition
and the second one
 come from  (\ref{difference-operator_sym}) and (\ref{M_cond}), respectively.
The cyclic transformation is given by the following cyclic permutation, 
\begin{eqnarray}
w \rightarrow z, \qquad z \rightarrow \frac{1}{wz}, \qquad \frac{1}{wz} \rightarrow w,
\label{cyc_w_rep}
\end{eqnarray}
and, of course, (\ref{CLR_w_rep}) is invariant under this permutation.

The CLR itself has no information on the normalizations of $M$ and $\Delta$.
To construct the solutions which are correctly normalized in the continuum limit, we have to require the further conditions,
\begin{eqnarray}
 &&\hM(1,1) =1, \label{cond_hM_2}\\
 &&\left.\frac{\partial\hD(w)}{\partial w}\right|_{w=1}=1. \label{cond_hD_2}
\end{eqnarray}
The first condition is the normalization of the cubic interaction term. The second one guarantees that $\D_{mn}$ becomes  the differential operator of the correct degree one in the continuum limit, that is, $\frac{d}{dt}$.

Once the momentum representation of a function is obtained, we may extend the domain of the function  by an analytic continuation.
As the domain of $\Delta(\omega)$, we can take the annulus,% ${\cal D}_K$,
 \begin{eqnarray}
{\cal D}_{K} = \{ \, w \in {\mathbb C}\, | \,  e^{-K} < |w| < e^K \},
\end{eqnarray}
for a positive constant $K$, because $\Delta(\omega)$ is holomorphic on ${\cal D}_K$ if the corresponding $d_l$ satisfies (\ref{local_d}). 
On the other hand, if the cyclic symmetry is taken into account, ${\cal D}_K \times {\cal D}_K$ is not so natural 
as the domain of $M(w,z)$ satisfying  the CLR (\ref{CLR_w_rep}),
because  the space is not closed under the cyclic transformation (\ref{cyc_w_rep}).

Instead, a natural choice of the domain is %the following ${\cal C}_{K}$ defined by 
\begin{eqnarray}
{\cal C}_{K} = \{ \, (w,z) \in {\mathbb C}^2 \, | \,  e^{-K} < s < e^K, \ {\rm for \, } s=|w|,|z|,|wz| \, \},
\label{cyclic_annulus}
\end{eqnarray}
for a positive constant $K$. This domain is different from ${\cal D}_K \times {\cal D}_K$ due to the  condition on the product $wz$.  
In fact, we find inclusion relations,
\begin{eqnarray}
{\cal D}_{K/2} \times {\cal D}_{K/2} 
\ \ \subset  \ \ 
{\cal C}_K
\ \ \subset \ \ 
{\cal D}_{K} \times {\cal D}_{K}. 
\label{inc_annulus}
\end{eqnarray}
On  ${\cal C}_K$, the cyclic symmetry is consistently defined  and the CLR is also naturally defined as a closed equation.  Therefore, we call the domain (\ref{cyclic_annulus}) $cyclic$ $invariant$ $annulus$.

As shown in \cite{Kato-1},  the exponential locality condition for $S_{kl}$  (\ref{S_local}) is equivalent to the fact that the corresponding $M(w,z)$ is holomorphic on ${\cal D}_{K} \times {\cal D}_{K}$. 
A similar equivalence holds for the cyclic locality. The proof does not need  (\ref{cond_hM_1}), namely, can be shown 
for a general matrix $S_{kl}$ satisfying the locality conditions,  (\ref{cyc_local}) and 
\begin{eqnarray}
 \vert\vert S_{ -k, l} \vert\vert  \le C\exp \{ -K(k+l) \}, \qquad {\rm for} \ \  k,l \in {\mathbb N}.
 \label{cyc_local_2}
\end{eqnarray}
The two conditions (\ref{cyc_local}) and  (\ref{cyc_local_2})  are equivalent for the symmetric $S_{kl}$. 
%Then we require the cyclic locality condition for $\vert\vert S_{-k, l}\vert\vert$,
In addition, the corresponding $M$ is 
given by (\ref{hM}). 
The following lemma tells us that the cyclic locality is equivalent to the holomorphicity of $M$ on the cyclic invariant annulus. 

\vspace{2mm}
\begin{lem}
 The following two propositions are equivalent to each other:
 \begin{itemize}
  \item[1.] $S_{kl}$ satisfies the cyclic locality.
  \item[2.] The corresponding $\hM(w,z)$ is holomorphic on the cyclic invariant annulus. 
 \end{itemize}
 where the cyclic locality and the cyclic invariant annulus are given by (\ref{cyc_local}), (\ref{cyc_local_2}) and (\ref{cyclic_annulus}), respectively.
\end{lem}
\vspace{2mm}

%The cyclic locality and the cyclic invariant annulus are given by (\ref{cyc_local}) and (\ref{cyclic_annulus}), respectively.
%\vspace{2mm}

\noindent
{\it Proof:} \ 
First, 
let us consider the following    
$\hM(w,z)$,
\begin{eqnarray}
  \hM(w,z)
  &=& S_{00} + \sum_{k\in{\mathbb N}, \, l \in {\mathbb N}_+} %\frac{1}{N_{kl}}
              \left(S_{k,-l}\frac{w^{k}}{z^l}+S_{-k-l,-k}\frac{1}{w^{k+l}z^{k}}+S_{l,k+l}w^{l}z^{k+l} \right. \nonumber \\
&&           \hspace{5cm} \left.  +\ (k,l\rightarrow -k,-l) \frac{}{} \right),
\label{hM_cyc}
\end{eqnarray}
%where $N_{kl}=1$ ($k,l\in{\mathbb N_+}=\{1,2,3,\cdots\}$), $N_{0k}=N_{k0}=2$ ($k\in{\mathbb N_+}$) and $N_{00}=6$. 
%
where ${\mathbb N}=\{0,1,2,\cdots\}$ and ${\mathbb N_+}=\{1,2,3,\cdots\}$. 
This $M$  equals to (\ref{hM}) if the right-hand side of (\ref{hM_cyc}) is convergent.
Note that the cyclically equivalence elements  (\ref{equiv_class_s}) gives the coefficients of (\ref{hM_cyc}).
Suppose that  the first proposition holds. 
Then, (\ref{hM_cyc}) is absolutely uniformly convergent on ${\cal C}_{K}$. Therefore, (\ref{hM})  is holomorphic  on ${\cal C}_K$. 
 
Conversely, 
suppose that the second proposition holds. Then, for $0 < {}^{\forall} K' <K$,
the inverse Fourier transformation,  
\begin{eqnarray}
 S_{kl} = 
  \oint_{|w|=1}\frac{dw}{2\pi i} \oint_{|z|=1}\frac{dz}{2\pi i}
          \ \hM(w,z)\  w^{-k-1} z^{-l-1},
\end{eqnarray}
gives 
\begin{eqnarray}
&& |S_{k,-l}| =
                \left|\oint_{|w|=e^{K'}}\frac{dw}{2\pi i} \oint_{|z|=e^{-K'}}\frac{dz}{2\pi i}
                     \ \hM(w,z)\ \frac{z^{l-1}}{w^{k+1}}\right|
          \le C e^{-K'(k+l)},
          \label{changed_S}
\end{eqnarray}
for $k,l \in {\mathbb N}$, where
$C$ is the maximum value of $|\hM|$ on ${\cal C}_K$.
We changed the integration contours in (\ref{changed_S}).  This is actually possible as explained below.
Let $w({\theta_1})=e^{i\theta_1}, z({\theta_2})=e^{i\theta_2}$ be the original contours for $\theta_1, \theta_2 \in [0,2\pi]$.  Then, for $t\in [0,1]$, 
the contours defined on ${\cal C}_K$,
\begin{eqnarray}   
 w({\theta_1},t) = w({\theta_1})e^{K't}, \qquad z({\theta_2},t) = z({\theta_2})e^{-K't}.
\end{eqnarray}
gives the original ones at $t=0$ and the contours of (\ref{changed_S}) at $t=1$.
%\begin{eqnarray}
% |w| = e^{K'}, \qquad |z| = e^{-K'}.
%\end{eqnarray}
%Therefore, the change of contours is justified. \hspace{2mm}  $\square$
%
Similarly,  
\begin{eqnarray}
 && |S_{-k-l,-k}| =
                \left|\oint_{|w|=e^{-K'}}\frac{dw}{2\pi i} \oint_{|z|=1}\frac{dz}{2\pi i}
                     \ \hM(w,z)\ (wz)^{k-1}w^{l}\right|
          \le C e^{-K'(k+l)}, \\
 && |S_{l,k+l}| =
                \left|\oint_{|w|=1}\frac{dw}{2\pi i} \oint_{|z|=e^{K'}}\frac{dz}{2\pi i}
                      \ \hM(w,z)\ (wz)^{-l-1}z^{-k}\right|
          \le C e^{-K'(k+l)},
\end{eqnarray}
for $k,l \in {\mathbb N}$. 
The same results for the remaining three elements can be obtained by the replacement, $k,l\rightarrow -k,-l$. 
Therefore, the first proposition (\ref{cyc_local}) and  (\ref{cyc_local_2}) holds. \qquad ${\Box}$%\qed
\vspace{5mm}

%%%%%%%%%%%%%%%%%%%%%%%%%%%%%%%%%%%%%%%%%%%%%%%%%%%%%%%%%%%%%%%%%%%%
%                                                                                                   %
%                                                                                                   %
%                                                                                                   %
%          Solution space and degree associated with cyclic symmetry             % 
%                                                                                                   %
%                                                                                                   % 
%                                                                                                   %
%%%%%%%%%%%%%%%%%%%%%%%%%%%%%%%%%%%%%%%%%%%%%%%%%%%%%%%%%%%%%%%%%%%%
\vspace{5mm}
\subsection{Solution space and degree associated with cyclic symmetry}

In the previous subsection, we saw the momentum representation of the CLR. All of the functions are  written as holomorphic functions on the cyclic invariant annulus ${\cal C}_K$ (\ref{cyclic_annulus}).   
For later use, we define two important sets of the holomorphic functions:
a space of the cyclic invariant holomorphic functions and 
a solution space of the CLR.
Moreover,  we introduce a new type of degree compatible with the cyclic symmetry.

Let ${\cal D}$ be a domain in two complex plains $(w,z) \in {\mathbb C}^2$, and let ${\cal {H}}({\cal D})$ be a set of two variable  holomorphic functions in ${\cal D}$.  In particular, 
if ${\cal D}$ is a domain that includes any pairs $(w,z)$ and $(z,w)$, 
then we can divide the space ${\cal {H}}({\cal D})$ into two sectors,\begin{eqnarray}
{\cal H}^\pm ({\cal D}) = \{ \,f \, | \ f \in {\cal{ H}}(D), \ \ f(w,z)= \pm f(z,w), \ {\rm for \ } w,z \in {\cal D} \, \},
\end{eqnarray}
%where ${\cal {H}}({\cal D})$ is a set of holomprphic functions on ${\cal D}$. 
 which are symmetric$(+)$ and asymmetric$(-)$ under the interchange of $w$ and $z$.
%${\cal H} $ is a direct sum of ${\cal H}^\pm$. 
%

Let us introduce the spaces which are spanned by the cyclic invariant holomorphic functions, 
\begin{eqnarray}
 {\cal H}^\pm_{\rm cyc} ({\cal C}_K)
  =
 \{ \,
  f \, | \, f \in {\cal H}^\pm ({\cal C}_{K} ) {\rm \  is \ cyclic \ invariant}. \}.
  \label{H_pm}
\end{eqnarray} 
The full space ${\cal H}_{\rm cyc}$ is the direct sum of ${\cal H}^+_{\rm cyc}$ and ${\cal H}^-_{\rm cyc}$. 
The easiest way to find elements of ${\cal H}^{\pm}_{\rm cyc} $ is to use 
the following two Laurent polynomials $\rho_+$ and $\rho_-$,
\begin{eqnarray} 
 \rho_+(w,z) \equiv w+z+\iwz, 
 \label{rho+}\\
 \rho_-(w,z) \equiv \iw+\iz+wz,
 \label{rho-}
\end{eqnarray}
and a special combination,
 \begin{eqnarray}
\iota(w,z) =(w-z)\left(z-\frac{1}{wz}\right)\left(\frac{1}{wz}-w\right).
\label{iota}
\end{eqnarray}  
They are manifestly cyclic invariant,  and $\rho_\pm$ are $w,z$-symmetric while $\iota$ is $w,z$-asymmetric. 
Suppose $g(x,y)$ is a two variable holomorphic function of $x,y \in {\mathbb C}$.
Then, $g(\rho_+ ,\rho_-)$ is in $ {\cal H}^+_{\rm cyc}$ and 
$iota$  (\ref{iota})$\times g(\rho_+ ,\rho_-)$ is in  ${\cal H}^-_{\rm cyc}$.

The solution space of the CLR is given by
%or a difference operator $\Delta$, let us define a solution space of the CLR on the cyclic invariant annulus ${\cal C}_K$ as follows:
\begin{eqnarray}
 {\cal M}^\pm ({\Delta},{\cal C}_K)
 =
 \{ \,
  M \, | \, M \in {\cal H}^\pm ({\cal C}_{K} ) {\rm \ satisfies \ the \ CLR \ with\ } \Delta. \}, 
  \label{M_pm}
\end{eqnarray}
where $\Delta$ is a difference operator. 
From the condition (\ref{cond_hM_1}),  this paper is mainly concerned with the solution in the plus sector.
As explained below, ${\cal M}^{\pm}$ have infinite elements. 
 A holomorphic function,
\begin{eqnarray}
M^-(w,z) = \Delta(\omega)- \Delta(z),
\end{eqnarray}
 is an element of $ {\cal M}^-(\Delta,{\cal C}_K)$.
 We can also find an element of the plus sector, 
 \begin{eqnarray}
 M^+(w,z) = \iota(w,z) M^-(w,z). % \ \ \in \  {\cal M}^+(\Delta,{\cal C}_K).
 \end{eqnarray}
Therefore, both the spaces ${\cal M}^\pm$ are not empty for any difference operator. 
%
%
%A simple question is how many elements 
%${\cal M}^\pm$ have, except for the ambiguities of the overall and relative constants.
Once a solution $M \in  {\cal M}^{\pm} ({\Delta},{\cal C}_K)$ is found, 
% 
%Suppose that $M \in  {\cal M}^+({\Delta},{\cal C}_K)$.  
we can give another solution,
\begin{eqnarray}
 \hM^{\prime}(w,z) \equiv f(w,z) \hM(w,z),\qquad 
 {\rm for\ }{}^\forall f \in  {\cal H}^+_{\rm cyc} ({\cal C}_K).
\end{eqnarray}
Since  $f$  is arbitrary,  $ {\cal M}^\pm$ have infinite elements except for the ambiguities of the overall and the relative constants.

Some concept  is  required to classify the solutions because there are infinite number of solutions. 
For polynomial-type solutions, some degree will be useful.
In particular, if the cyclic symmetry is taken into account, it should be compatible with the symmetry. 
%we will define the degree consistent  with the cyclic symmetry.
%
As the first trial, let us take $|k|+|l|$ as the degree for a monomial $w^kz^l$ $(k,l\in{\mathbb Z})$.
It is immediately clear that $\rho_\pm$ consist of the three monomials of different degree and in other words the naive definition of the degree is not compatible with the cyclic symmetry. In order to coexist with the cyclic symmetry, $w, z$ and $\frac{1}{wz}$ should have the same degree.
The reasonable degree is defined as follows:
%Let us introduce the degree compatible with the cyclic symmetry as follows: 
%
\begin{dfn} The degree of a Laurent monomial $w^kz^l\ (k,l\in{\mathbb Z})$ is defined as
\begin{eqnarray}
 \renewcommand{\arraystretch}{1.2}
     {\rm degree\ of\ } w^kz^l &=&
 \left\{\begin{array}{ll}
     \max\{|k|,|l|\}&\quad {\rm for\ } k,l {\rm \ have \ the \ same \ sign}, \\
     |k|+|l|&\quad {\rm for\ the \ others }.
      \\ \end{array}\right.
 \label{order}
\end{eqnarray}
\end{dfn}
\renewcommand{\arraystretch}{1}
This definition and the cyclic symmetry fit well together because
all of $w,z$ and $\iwz$ have degree one. More generally, the cyclically equivalent elements (\ref{equiv_class_s})
correspond to  the monomials of the same degree. 
Unless otherwise noted, the degree of the monomials which will be used in  the following 
 means (\ref{order}). 

Furthermore, the degree of a Laurent polynomial is defined as the highest degree of the monomials in it. Then, both $\rho_+$ (\ref{rho+}) and $\rho_-$ (\ref{rho-}) are polynomials of degree one.
The degree of any polynomial multiplied by each of $w, z$ and $\frac{1}{wz}$ does not always increase by one. In contrast,  it is easy to show that the degree of any polynomial multiplied  by $\rho_\pm$ increases by one.

%%%%%%%%%%%%%%%%%%%%%%%%%%%%%%%%%%%%%%%%%%%%%%%%%%%%%%%%%%%%%%%%%%%%
%                                                                                                   %
%                                                                                                   %
%                                                                                                   %
%               An expansion by cyclic invariant polynomials                           % 
%                                                                                                   %
%                                                                                                   % 
%                                                                                                   %
%%%%%%%%%%%%%%%%%%%%%%%%%%%%%%%%%%%%%%%%%%%%%%%%%%%%%%%%%%%%%%%%%%%%
\vspace{5mm}
\subsection{Expansion by cyclic invariant Laurent polynomials }
As well-known, the holomorphic functions can be expanded as the Laurent series  on an annulus.
Similarly, cyclic invariant holomorphic functions can be expanded   by  cyclic invariant Laurent polynomials on a cyclic invariant annulus.  
In this section,  we give a proof of the expansion. The expansion is used to formulate a recurrence relation of the CLR in the next subsection.

\begin{thm}
Let $K$ be a positive constant and let $f\in {\cal H}_{\rm cyc}({\cal C}_K)$. 
Then, it can be uniquely expressed as the following series,
 \begin{eqnarray}  
 f(w,z) = c + \sum_{n=1}^\infty \sum_{i=-n+1}^n \, c_{ni} p_{ni}(w,z),
 \label{cyclic_Laurent}
  \end{eqnarray}  
where the complex constants $c_{ni}$ are given by
\begin{eqnarray}  
c_{ij} = c_{j,-i} =\oint_{|w|=1} \frac{dw}{2 \pi i} \oint_{|z|=1} \frac{dz}{2 \pi i} \, f (w,z) w^{i-1} z^{j-1},  \qquad  {\rm for} \  i,j\ge 0,
% && f_+(w,z) =f(w,z), \quad  f_- (w,z) =f(z,w). 
%&& c_{n, -i} = \oint_{|w|=1} \frac{dw}{2 \pi i} \oint_{|z|=1} \frac{dz}{2 \pi i} \, f(w,z) w^{i-1} z^{n-1},
% \quad \ \ (i=0,1,\cdots,n), 
 \label{cyclic_Laurent_coef}
 \end{eqnarray}  
%
%for $ f_+(w,z) =f(w,z), \ f_- (w,z) =f(z,w)$, 
with $c=c_{00}$,
and the cyclic invariant Laurent polynomials $p_{ni} (n \in {\mathbb N}, |i| \le n )$ are defined by
\begin{eqnarray}  
p_{ni}(w,z) =p_{n,-i}(z,w)=
  \frac{w^i}{z^{n-i}} + \frac{1}{w^nz^i} + w^{n-i} z^n
, %\quad (i=0,1,\cdots,n).
\label{def_p_ni}
\end{eqnarray}
for $i=0,\cdots,n$.
The right-hand side of (\ref{cyclic_Laurent}) is 
absolutely uniformly convergent on any compact subset of ${\cal C}_K$.
\end{thm}
\vspace{2mm}

\noindent {\it Proof:}
Once $f$ is written as (\ref{cyclic_Laurent}), 
the uniqueness of the expression is mostly trivial because $f=0$ means $c_{ni}=0$.
We only have to show whether $f$ can be  expressed as the convergent series (\ref{cyclic_Laurent}) on ${\cal C}_K$. 

From (\ref{inc_annulus}), the cyclic invariant annulus ${\cal C}_K$ has an annulus $D_{K/2} \times D_{K/2}$
as a subset. 
On the annulus $D_{K/2} \times D_{K/2}$, $f$ can be expanded as the Laurent series,
\begin{eqnarray}
f(w,z) = \sum_{m=-\infty}^\infty \sum_{n=-\infty}^\infty \alpha_{mn} w^m z^n, 
\label{exp_f_ann}
\end{eqnarray}
where
\begin{eqnarray}  
 \alpha_{mn} = \oint_{|w|=1} \frac{dw}{2 \pi i} \oint_{|z|=1} \frac{dz}{2 \pi i} \, f(w,z) w^{-m-1} z^{-n-1},
 \label{Laurent_coef}
 \end{eqnarray}  
 and the right-hand side is uniformly absolutely-convergent on any compact subset of the annulus.

 The absolute convergence of the series (\ref{exp_f_ann}) allows us to change the order of the summation in (\ref{exp_f_ann}) and write $f$ as
 \begin{eqnarray}
 f(w,z)  
 &=& \alpha_{00} + \sum_{n=1}^\infty
\left\{
 \alpha_{0,-n} \frac{1}{z^{n}}
    +  \alpha_{-n0} \frac{1}{w^{n}}
    +  \alpha_{nn} w^{n} z^{n} 
    \right. \nn \\
&& \left. \hspace{1.6 cm} 
    +  \alpha_{n0} w^n
    +  \alpha_{-n,-n} \frac{1}{z^{n} w^n}
    +  \alpha_{0n} z^{n} \right\} \nn  \\
&& \hspace{0.7 cm} 
  +  \sum_{n=1}^\infty \sum_{i=1}^{n-1} \left\{
       \alpha_{i,-n+i} \frac{w^i}{z^{n-i}}
    +  \alpha_{-n,-i} \frac{1}{w^{n} z^i}
    +  \alpha_{n-i,n} w^{n-i} z^{n} 
    \right.
    \nn
    \\
&& 
\left.
\hspace{2.6 cm} 
    +   \alpha_{-n+i,i} \frac{z^i}{w^{n-i}}
    +  \alpha_{-i,-n} \frac{1}{z^{n} w^i}
    +  \alpha_{n,n-i} z^{n-i} w^{n} 
 \right\}.
 \end{eqnarray} 
% where $N_{nn}=N_{n0}=2$ and the others are unity.
 %
 %
 The cyclic symmetry of $f$ tells us that  the coefficients $\alpha_{mn}$ satisfy
 \begin{eqnarray}  
 && \alpha_{i,-n+i}=\alpha_{n-i,n}=\alpha_{-n,-i}  \equiv c_{ni}, \\
 &&\alpha_{-n+i,i}=\alpha_{n,n-i}=\alpha_{-i,-n} \equiv c_{n,-i}.
 \label{obtained_coef}
 \end{eqnarray}  
 Therefore, 
\begin{eqnarray}
 f(w,z)   = c_{00} + \sum_{n=1}^\infty \sum_{i=-n+1}^n \, c_{ni} p_{ni}(w,z).
 \label{cyc_Laurant_expansion}
\end{eqnarray}

The remaining task is to show  that the right-hand side of $(\ref{cyc_Laurant_expansion})$ is well-defined on $C_K$.
Now, let us take any compact subset $U$ in ${\cal C}_{K}$. For appropriate $K' (0<K'<K)$, $U$ is included in ${\cal C}_{K'}$.
Then,  by changing the integral contours of $\alpha_{mn}$ (\ref{Laurent_coef}) as done in the proof of the lemma in section \ref{sec:cyclic_annulus}, we can show that
 \begin{eqnarray}
|c_{ni}|  \le C e^{-K' n},
\label{cond_local_c_ni}
 \end{eqnarray}
 where $C$ is some positive constant, and for ${}^\forall (w,z) \in U$, 
 \begin{eqnarray}
 |p_{ni}(w,z)| < 3e^{K'n}.
\label{cond_local_p_ni}
 \end{eqnarray}
From (\ref{cond_local_c_ni}) and (\ref{cond_local_p_ni}), the right-hand side of (\ref{cyc_Laurant_expansion}) is absolutely uniformly convergent on any compact subset of ${\cal C}_{K}$, and the theorem holds. \qquad $\Box$
%\qed
\vspace{5mm}
 
From the definition (\ref{def_p_ni}), each 
$p_{ni} \in {\cal H}_{\rm cyc}({\cal C}_K)$ consists of the three $n$-th degree monomials. $p_{nn}(w,z)=p_{n,-n}(w,z)$ and  two  $p_{n0}$ come from $p_{n,+i}$ and $p_{n,-i}$ are the same.
 All of $p_{ni} (i=-n+1,\cdots,n)$ are independent of each other.
Thus, the expansion (\ref{cyclic_Laurent}) can be regarded as an expansion by the cyclic invariant Laurent polynomials,
which are independent of each other,
  with respect to the degree (\ref{order}).

 The cyclic invariant Laurent polynomials can be divided into two sectors,
% Let us define 
 \begin{eqnarray}
 p^\pm_{ni}(w,z) =\frac{1}{N_{ni}} \left( p_{ni}(w,z) \pm p_{ni}(z,w) \right) \ \in \  H^\pm_{\rm cyc}({\cal C}_K),
 \end{eqnarray}
where $N_{nn}=N_{n0}=2$ and the other $N_{ni}$ are unity: 
 \begin{eqnarray}
p^\pm_{ni}(w,z)  = \frac{w^i}{z^{n-i}} + \frac{1}{w^nz^i} + w^{n-i} z^n
\pm (w \leftrightarrow z),
\label{p_pm}
 \end{eqnarray}
 for $i=1,2,\cdots,n-1$,
and  %for $i=0,n$, 
 \begin{eqnarray}
&& p^+_{n0}(w,z)  = \frac{1}{w^n} + \frac{1}{z^n} +w^n z^n, 
\qquad\quad p^-_{n0}(w,z) =0,
\label{p_plus1}
\\
&& p^+_{nn}(w,z)  = w^n+ z^n + \frac{1}{w^n z^n},
\qquad\quad p^-_{nn}(w,z) =0.
\label{p_plus2}
 \end{eqnarray}
Note that $p_{n0}$ and $p_{nn}$ are invariant under the interchange of $w$ and $z$.
 
 The theorem gives the expansion for the general cyclic invariant Laurent polynomials included in ${\cal H}_{\rm cyc}$.  
 Similar expansions hold for both the sectors $H^\pm_{\rm cyc}$. We only give the results without the proofs because they are mostly the same as the proof of the theorem (\ref{cyclic_Laurent}).

 \vspace{2mm}

 \begin{cor_1}

 Let $K$ be a positive constant and let $f \in H^+_{\rm cyc}({\cal C}_K)$. Then, it can be uniquely expressed as the following series,
 \begin{eqnarray}  
 f(w,z) = c + \sum_{n=1}^\infty \sum_{i=0}^n \, c_{ni} p^+_{ni}(w,z),
 \label{cyclic_Laurent_plus}
  \end{eqnarray}  
where the complex constants $c_{ni} \, (n \in {\mathbb N}, i=0,\cdots,n )$ are given by
\begin{eqnarray}  
 c_{ni} = \oint_{|w|=1} \frac{dw}{2 \pi i} \oint_{|z|=1} \frac{dz}{2 \pi i} \, f(w,z) w^{n-1} z^{i-1},
 \label{cyclic_Laurent_plus_coef}
 \end{eqnarray} 
 with $c=c_{00}$, 
and the cyclic invariant Laurent polynomials $p^+_{ni} \, (n \in {\mathbb N},  i=0,\cdots,n)$ are given by 
  (\ref{p_pm}), (\ref{p_plus1}) and (\ref{p_plus2}).
The right-hand side of (\ref{cyclic_Laurent_plus}) is 
absolutely uniformly convergent on any compact subset of ${\cal C}_K$.
 \end{cor_1}
 \vspace{2mm}

 \begin{cor_2}
 Let  $K$ be a positive constant and let $f \in H^-_{\rm cyc}({\cal C}_K)$. Then, it can be uniquely expressed as the following series,
 \begin{eqnarray}  
 f(w,z) = \sum_{n=2}^\infty \sum_{i=1}^{n-1} \, c_{ni} p^-_{ni}(w,z),
 \label{cyclic_Laurent_minus}
  \end{eqnarray}  
where the complex constants $c_{ni} \, (n \ge 2, \  i=1,\cdots,n-1 )$ are given by
\begin{eqnarray}  
 c_{ni} = \oint_{|w|=1} \frac{dw}{2 \pi i} \oint_{|z|=1} \frac{dz}{2 \pi i} \, f(w,z) w^{n-1} z^{i-1},
 \label{cyclic_Laurent_minus_coef}
 \end{eqnarray}  
and the cyclic invariant Laurent polynomials $p^-_{ni} \  (n \ge 2, \  i=1,\cdots,n-1)$ are given by (\ref{p_pm}).
The right-hand side of (\ref{cyclic_Laurent_minus}) is 
absolutely uniformly convergent on any compact subset of ${\cal C}_K$.
 \end{cor_2}

.
%%%%%%%%%%%%%%%%%%%%%%%%%%%%%%%%%%%%%%%%%%%%%%%%%%%%%%%%%%%%%%%%%%%%
%                                                                                                   %
%                                                                                                   %
%                                                                                                   %
%               Recurrene relations and fundamental solutions                        % 
%                                                                                                   %
%                                                                                                   % 
%                                                                                                   %
%%%%%%%%%%%%%%%%%%%%%%%%%%%%%%%%%%%%%%%%%%%%%%%%%%%%%%%%%%%%%%%%%%%%
\vspace{5mm}
\subsection{Recurrence relation and fundamental solutions}
The recurrence relation for the coefficients $S_{kl}$ in (\ref{hM}) is useful to  construct the finite-dimensional solutions. 
We derive the relation from the CLR (\ref{CLR_w_rep}) and construct the solutions of lower degree for the naive symmetric difference operator.

By extracting the coefficients of the monomials $w^k z^l$ from the CLR, one can formally obtain the recurrence relation. But the obtained set of the equations are multiply degenerate, due to the cyclic symmetry. 
To avoid the degeneracy, we can use the expansion by the cyclic invariant  Laurent polynomials, which gives the set of the independent equations.
Using (\ref{cyclic_Laurent_plus}), 
for $n \ge 0,\, i=0,\cdots, n$,
let us define the set, 
\begin{eqnarray}
  A_{ni} \equiv \oint_{|w|=1} \frac{dw}{2 \pi i} \oint_{|z|=1} \frac{dz}{2 \pi i} \ w^{n-1} z^{i-1} \times ({ \rm l.h.s. \  of \  the\  CLR} ).
\end{eqnarray}
From the residue theorem,  
\begin{eqnarray} 
 A_{ni} \equiv \sum_{m\in{\mathbb Z}} d_m(S_{i+m,-n+i+m}  +  S_{-n+m,-i+m} + S_{n-i+m, n+m}).
 \label{Amn_general}
\end{eqnarray}
The locality conditions  (\ref{cyc_local}) and (\ref{local_d}) guarantee that 
(\ref{Amn_general}) is well-defined because the right-hand side is exponentially suppressed at  large $m$.
The recurrence relation is, therefore, given by
\begin{eqnarray}
 A_{ni} = 0,\qquad {\rm for\ } \ n\in{\mathbb N}, \ i=0,1,\cdots,n. \label{Amn_0}
\end{eqnarray}
By solving the recurrence relation (\ref{Amn_0}), we can find solutions of the CLR.

In the case of the naive symmetric difference operator (\ref{sol_D}), 
\begin{eqnarray}
 \hD_s(w)=\frac{1}{2}\left(w-\iw\right),
 \label{D_naive}
\end{eqnarray}
the corresponding recurrence relation (\ref{Amn_general}) is 
\begin{eqnarray}
 2A_{ni}&=&S_{i+1,-n+i+1}+S_{-n+1,-i+1}+S_{n-i+1,n+1}\nn\\
       && -S_{i-1,-n+i-1} - S_{-n-1,-i-1} - S_{n-i-1,n-1}.\label{Amn}
\end{eqnarray}
%
%where $k, l \in{\mathbb N}$. 
%$A_{00}$ relates only monomials of degree one, and $A_{kl}\ (k, l\neq0)$ relate three monomials with degrees $k+l-1$, $k+l$ and $k+l+1$.
%Therefore the monomials of degree $n$ are related to the recurrence relations $A_{kl}=0$ for $n-1\leq k+l \leq n+1$, 
The monomials of degree $n$ correspond to the $A_{mi}=0$ for $n-1\leq m \leq n+1$ so that the general solution of degree $n$ can be obtained by solving $A_{mi}=0$ for $m \leq n+1$.\footnote{
%Let us count the number of parameters 
For the $n$-th degree polynomial solution, 
%After taking $S_{kl}=0$ associated with the monomials of degree  higher than $n$,  
the number of independent $S_{kl}$ is $\frac{(3n+2)(n+1)}{2}$. Naively, the number of independent equations $A_{mi}$ (\ref{Amn}) is $\frac{(n+2)(n+3)}{2}$,  
however it is found that there are two following relations among $A_{mi}$, 
\begin{eqnarray}
&& A_{00}
 +3\sum_{k=1}^{\left[\frac{n+1}{2}\right]}
  \left(
   A_{2k,0}+A_{2k,2k}+2\sum_{l=1}^{k-1}A_{2k,2l}
  \right)
 =0,\\
&& 2\sum_{k=1}^{\left[\frac{n+1}{2}\right]} \sum_{l=0}^{k-1} A_{2k, 2l+1}%\nn\\&&
 %\quad
 +\sum_{k=0}^{\left[\frac{n}{2}\right]}
  \left(
   A_{2k+1,0}+A_{2k+1,2k+1}+2\sum_{l=1}^{k}(A_{2k, l}+A_{2k+1, 2k-l+1})
  \right)
 =0.
\end{eqnarray}
Therefore, the number of independent equations is less than or equal to $\frac{(n+2)(n+3)}{2}-2$.
Fro this consideration, the $n$-th degree general solution has $n^2$ parameters at most. In appendices A and B, it will be found that the $n$-th degree general solution can be given by $n^2$ parameters.
}

% Let us solve the recurrence relations $(\ref{Amn})$ degree by degree from one.
Let $\hM_{n}$ be the general solution of degree $n$.
By solving $A_{ni}=0\ (n\leq 2)$, % according to the consideration above, 
we find
\begin{eqnarray}
 \hM_1(w,z) = a\left(w+z+\iw+\iz+2 \right),
 \label{hM_1}
\end{eqnarray}
where $a\in{\mathbb C}$, and $a=\frac{1}{6}$ for the normalized solution, (\ref{cond_hM_2}).
Similarly, %the general solution of degree two $\hM_2$ is %
\begin{eqnarray}
 \hM_2(w,z) &=& a\left(w+z+\iw+\iz+2\right) \nn\\&& 
              +b\left(2wz+\frac{2}{wz}+\frac{w}{z}+\frac{z}{w}\right) \nn\\&&
              +c\left(\frac{1}{w^2}+\frac{1}{z^2}+\iw+\iz+w^2z+wz^2\right)\nn\\&&
              +d\left(w^2+z^2+w+z+\frac{1}{w^2z}+\frac{1}{wz^2}\right),
 \label{hM_2}
\end{eqnarray}
where $a, b, c$ and $d \in {\mathbb C}$ and  $6(a+b+c+d) = 1$ for the normalized solution (\ref{cond_hM_2}). The first polynomial in the left-hand side of (\ref{hM_2}) is $\hM_1$. The other three solutions have degree two, and the third and the fourth polynomials can be rewritten as the linear combinations of the first and the second ones with $\rho_{\pm}$, (\ref{rho+}) and (\ref{rho-}).

In fact, it is possible to express $\hM_2$ as 
\begin{eqnarray}
 \hM_2
   = (a-c-d+c \rho_-+d \rho_+) B_1 
             + (b-c-d) B_2,
             \label{hM_2_gen}
\end{eqnarray}
where the two solutions  $B_1$ and $B_2$ are defined by
\begin{eqnarray}
 B_1(w,z) &\equiv& w+z+\iw+\iz+2,\label{B1}\\
 B_2(w,z) &\equiv& 2wz+\frac{2}{wz}+\frac{w}{z}+\frac{z}{w}.\label{B2}
\end{eqnarray}
%
%Here, we used $+$ for $B_i$ to say that they are the solutions of ${\cal M}^+(\Delta_s,{\cal C}_K)$, clearly.
Again, $B_1$ is $M_1$, and $B_2$ is the second polynomial in the left-hand side of (\ref{hM_2}).
They are invariant under not only $w\leftrightarrow z$, but also $w,\ z \leftrightarrow \iw,\ \iz$, 
and hereafter will be referred to as fundamental solutions.

Further direct investigation for the general solution of degree six (see appendix A) tells us  that the solution can be  also expressed as a linear combination of $B_1$ (\ref{B1}) and $B_2$ ($\ref{B2}$) with coefficients which are polynomials of $\rho_{\pm}$ such as $(\ref{hM_2_gen})$.  
As a natural extension, it would be expected that the general solution   
is written in terms of $B_1$ and $B_2$ with holomorphic functions of $w,z$ which are invariant under the cyclic symmetry. 
We will prove this speculation as  theorem A in section 4.% by using delta operators introduced in the next subsection. 

%%%%%%%%%%%%%%%%%%%%%%%%%%%%%%%%%%%%%%%%%%%%%%%%%%%%%%%%%%%%%%%%%%%%
%                                                                                                   %
%                                                                                                   %
%                                                                                                   %
%                               Delta operators                                                % 
%                                                                                                   %
%                                                                                                   % 
%                                                                                                   %
%%%%%%%%%%%%%%%%%%%%%%%%%%%%%%%%%%%%%%%%%%%%%%%%%%%%%%%%%%%%%%%%%%%%
\vspace{5mm}
\subsection{Delta operators}
In the previous subsection, it was expected that any solutions of the CLR are given by $M=\sum_{i=1,2}f_{i}B_{i}$,
where $B_{i}$ are the two fundamental solutions, and $f_{i}$ are holomorphic functions of $w, z$ that are invariant under both the cyclic transformation and the interchange of $w$ and $z$. 
This expectation will be shown for the naive symmetric difference operator $\Delta_s$ (\ref{D_naive}) in section 4. In this subsection, we introduce mappings which play an important role in the proof.

Let us consider two mappings, referred to as delta operators,\footnote{
Let $\delta$ be a mapping defined by $\delta=(\delta_1,\delta_2)$,
\begin{eqnarray}
\begin{array}{cccc}
\delta :   &   {\cal M}^+(\Delta_s, {\cal C}_{K} )             &   \longrightarrow    &     {\cal H}^+_{\rm cyc}({\cal C}_K) \times    {\cal H}^+_{\rm cyc}({\cal C}_K).        \vspace{2mm}  \\[-4pt] 
              &   \rotatebox{90}{$\in$}                        &                   & \rotatebox{90}{$\in$}                        \vspace{2mm} \\[-4pt]       
              &  M                     & \longmapsto      & (f_1, f_2)                                     
\end{array}
\end{eqnarray}
Theorem A given in section 4 tells us that $\delta$ is a one-to-one and onto mapping.
}
\hspace{2cm} 
\begin{eqnarray}
\begin{array}{ccccc}
\delta_i :   &   {\cal M}^+(\Delta_s, {\cal C}_{K} )             &   \longrightarrow    &     {\cal H}^+_{\rm cyc}({\cal C}_K),           & \qquad  (i=1,2),   \vspace{2mm}  \\[-4pt] 
              &   \rotatebox{90}{$\in$}                        &                   & \rotatebox{90}{$\in$}                    &    \vspace{2mm} \\[-4pt]       
              &  M                     & \longmapsto      & f_i                                     &
\end{array}
\end{eqnarray}
for  the naive  symmetric difference operator $\Delta_s$, where 
$f_i$ are uniquely determined from $M$ through
\begin{eqnarray}
&& B_{i}\left(w,\frac{1}{wz}\right)M\left(z,\frac{1}{wz}\right)
  -(w \leftrightarrow z)
          %-B_{i}\left(z,\frac{1}{wz}\right)M\left(w,\frac{1}{wz}\right) \nonumber \\
%&& \hspace{5cm}
 \  = \ -  \iota(w,z) \, 2 \Delta_s \left(\frac{1}{wz}\right)  \epsilon_{ij} f_{j} (w,z), 
\label{delta_identity}    
\end{eqnarray}
where %$B_{i}$ are two fundamental solutions, and
$\epsilon_{12}=-\epsilon_{21}=1$ and $\epsilon_{11}=\epsilon_{22}=0$ and $\iota$ is given in (\ref{iota}).

Once the identity (\ref{delta_identity}) holds, the delta operators satisfy
\begin{eqnarray}
 && \delta_{i} (aM+bN) = a\delta_{i}M +b\delta_{i} N ,\label{delta_lin}\\
 && \delta_{i} B_{j} = \delta_{ij}, \label{delta_B}
\end{eqnarray}
where $M,N\in {\cal M}^+({\Delta_s},{\cal C}_K)$, $a,b\in {\cal H}^+_{\rm cyc} ({\cal C}_K)$ and $\delta_{ij}$ is the Kronecker delta.
The first equation (\ref{delta_lin}) means that $\delta_i$ are linear operators.
The second one (\ref{delta_B}) follows from  
\begin{eqnarray}
\epsilon_{ij} B_i \left(w,\frac{1}{wz} \right) B_j \left(z,\frac{1}{wz} \right) = - \iota(w,z) \, 2\Delta_s \left( \frac{1}{wz} \right).
\label{cyclic_Euler}
\end{eqnarray}
%
%
%
%Using equations above, we can show that
%\begin{eqnarray}
%\delta_{i}M=f_{i}, 
%\qquad {\rm for} \ \ M=\sum_{i=1}^{2}f_{i}B_{i}, \quad f_i \in {\cal H}_{\rm cyc} ({\cal C}_K),
%\end{eqnarray} 
%which plays an important role in a proof of main theorems.
%
%
In the remaining part of this subsection,  we show that the identity  (\ref{delta_identity}) holds and 
the resultant $f_{i}$ is a unique element of $ {\cal H}^+_{\rm cyc} ({\cal C}_K) $.

First, we define the following holomorphic functions, 
\begin{eqnarray}
\Phi(w,z) &=&
{\iota(w,z) \, 2 
 \Delta_s (w)
  \Delta_s (z)
\Delta_s \left(\frac{1}{wz}\right)
\ \in \ {\cal H}^-_{\rm cyc} ({\cal C}_K),
}  \label{Phi} \\
{\cal I}_i(w,z) &=&  \Delta_s(w)\Delta_s(z)\nonumber\\
&
\times&\epsilon_{ij}
\left[B_{j}\left(w,\frac{1}{wz}\right)M\left(z,\frac{1}{wz}\right)
      -B_{j}\left(z,\frac{1}{wz}\right)M \left(w,\frac{1}{wz} \right)\right],%\ (i=1,2),
      \label{I_function}
%\end{eqnarray}
\end{eqnarray}
Using the CLR (\ref{CLR_w_rep}), 
we find 
\begin{eqnarray}
{\cal I}_{i}, \ \in \ {\cal H}^-_{\rm cyc} ({\cal C}_K),
\label{I_i_Phi} 
\end{eqnarray}
and the set
\begin{eqnarray}
%{\cal I}_i(w,z) =0, \quad {\rm on} \ \ 
 {\cal Z}_K = \left\{(w,z)\in  {\cal C}_K | \, \Phi(w,z)=0 \right\},
\label{zero_of_I}
\end{eqnarray}
gives ${\cal I}_i(w,z) =0$.

Now, let us consider the following ratios,
\begin{eqnarray}
f_i(w,z) = \frac{{\cal I}_i(w,z)}{\Phi(w,z)},  \quad \ (i=1,2).
  \label{f_by_I}
\end{eqnarray}
%\begin{eqnarray}
%f_i(w,z) = \epsilon_{ij} \left(
%\frac{ {\cal I}_j(w,z)}
%{\iota(w,z) \, 2 
% \Delta_s (w)
%  \Delta_s (z)
%\Delta_s \left(\frac{1}{wz}\right)
%}
%\right),  \quad \ (i=1,2),
%  \label{f_by_I}
%\end{eqnarray}
Then, $f_i$ are the unique holomorphic functions in ${\cal C}_K- {\cal Z}_K$. 
The numerators of (\ref{f_by_I}) vanish at all of the points ${}^\forall (w,z) \in {\cal Z}_K$ where the denominator vanishes. 
Thus, we find that ${\cal Z}_K$ are removable singularities of $f_i$.
From this result with (\ref{Phi}) and (\ref{I_i_Phi}), 
it is easy to show  $f_i \in  {\cal H}^+_{\rm cyc}({\cal C}_K)$.

Moreover, using (\ref{I_function}) with (\ref{Phi}), 
$f_i$ can be also uniquely expressed as 
\begin{eqnarray}
f_i(w,z) = \epsilon_{ij} \left( \frac{B_{j}\left(w,\frac{1}{wz}\right)M\left(z,\frac{1}{wz}\right)
  -(w \leftrightarrow z)}{\iota(w,z) \, 2 \Delta_s \left(\frac{1}{wz}\right)} \right).
  \label{f_delta}
\end{eqnarray}
This shows that the delta mappings are well-defined.

%%%%%%%%%%%%%%%%%%%%%%%%%%%%%%%%%%%%%%%%%%%%%%%%%%%%%%%%%%%%%%%%%%%%
%                                                                                                   %
%                                                                                                   %
%                                                                                                   %
%                               Minimal solutions                                              % 
%                                                                                                   %
%                                                                                                   % 
%                                                                                                   %
%%%%%%%%%%%%%%%%%%%%%%%%%%%%%%%%%%%%%%%%%%%%%%%%%%%%%%%%%%%%%%%%%%%%
\vspace{5mm}
\subsection{Minimal solutions}
Let us consider particular solutions of the CLR for the naive symmetric difference operator. 
The solutions consist of the minimum number of terms (six terms with the same weight\footnote{ For instance, (\ref{B1}) and (\ref{B2}) have six terms if one regards $2$ as $1+1$, $2wz$ as $wz+wz$, and so on.}), 
and therefore, will be referred to as minimal solutions in the following.
In theorem B in section 4, we show that the general solution can be written as a linear combination of the minimal solutions.

By examining the structure of the recurrence relation, we find that, for any $k, l$, $S_{kl}$ appears in two  different $A_{ni}$, 
and no other common elements are in those $A_{ni}$. If $S_{kl}\neq0$, 
the other elements should be non-zero to cancel it. The shortest solutions can be realized by three $S_{kl}$ that cancel each other out in the three $A_{ni}$  as follows:
\begin{eqnarray}
 &&A_{ni}\simeq S_{kl}\pm S_{k^{\prime}l^{\prime}}=0\\
 &&A_{n^{\prime}i^{\prime}}\simeq S_{k^{\prime}l^{\prime}}\pm S_{k^{\prime\prime}l^{\prime\prime}}=0\\
 &&A_{n^{\prime\prime}i^{\prime\prime}}\simeq S_{k^{\prime\prime}l^{\prime\prime}} \pm S_{kl}=0,
\end{eqnarray}
where $\pm$ depends on $k, l$ and  $\simeq$ means that the irrelevant terms are ignored. 
 
A brute force investigation tells us that there are $2n-1$ minimal solutions of degree $n$, $\hL_{ni}$ $(i\in{\mathbb Z}, |i|\leq n-1)$. 
They are divided into the two types in which the combinations of the terms are different,
\begin{eqnarray}
 \hL_{ni} = 
 {\renewcommand\arraystretch{1.5}
 \left\{\begin{array}{ll}
 \hL_{ni}^{(1)}, &\qquad  (i=-n+1,-n+3,\cdots,n-3,n-1),\\
 \hL_{ni}^{(2)}, &\qquad  (i=-n+2,-n+4,\cdots,n-4,n-2),
 \end{array}\right.
 }\label{sol_min}
\end{eqnarray}
where
\begin{eqnarray}
&& \hL_{n,n-1-2i}^{(1)}
  =          w^{n}z^{n-1-i}
             +\frac{w^{i}}{z^{n-1-i}}
             +\frac{1}{w^n z^i}
             +(w\leftrightarrow z),\ (i=0,\cdots,n-1),
            \label{min_1}\\
&& \hL_{n,n-2i}^{(2)}
  = w^{n-1}z^{n-i}
    +\frac{w^i}{z^{n-i}}
    +\frac{1}{w^{n-1} z^{i}}
                +(w\leftrightarrow z),\ \quad(i=1,\cdots,n-1).
                \label{min_2}
\end{eqnarray}
Both the types of the minimal solutions consist of the six monomials of degree $n$ and $n-1$. 
The index $i$ of $L_{ni}^{(1)}$ runs integers with opposite even-odd parity of $n$, %, where its absolute values is less than $n(n-1)$.  
while that of the second type  runs integers with the same even-odd parity as $n$.
% where its absolute value is less than $n-1$.
We often abbreviate the type index $a$ of $\hL_{ni}^{(a)}$ because it can be uniquely determined from $i$.
When $n$ is odd(even), the first(second) type has $i=0$.\footnote{
From the definition (\ref{min_1}), (\ref{min_2}),
\begin{eqnarray}
 \hL_{n,-i}\left({w},{z}\right)
 =
 \hL_{n,i}\left(\frac{1}{w},\frac{1}{z}\right).
\end{eqnarray}
The minimal solutions with $i=0$ are self-dual. 
}

Let us consider a recurrence relation for the minimal solutions.
Using $\rho_{\pm}$, 
the minimal solutions can be related to the lower ones,
\begin{eqnarray}
&& \hL_{n,i}=
  \rho_{\pm} \hL_{n-1,i\pm1} - \hL_{n-1,i\pm3} - \hL_{n-2,i},
  \label{rec_L}
\end{eqnarray}
for $n\geq2, |i|\leq n-3$ and  $i=\mp(n-1), \mp(n-2)$, 
(double-sign corresponds). 
Here, we used 
\begin{eqnarray}
&&\hL_{n-1,\pm n}\equiv \hL_{n,\pm(n-2)},\label{1}\\
&&\hL_{n-1,\pm(n-1)}    \equiv \hL_{n-1,\pm(n-2)},\label{2}
\end{eqnarray}
for simplicity. 
Note that the left-hand side of  (\ref{1}) has degree $n$ nevertheless its index is $n-1$, 
while (\ref{2}) maintains the degree. 
The two minimal solutions $\hL_{10}$ and $\hL_{20}$ do not appear in the left-hand side of  (\ref{rec_L}) and 
are the fundamental solutions,
\begin{eqnarray}
 \hL_{10} &=& \hB_1,\\
 \hL_{20} &=& \hB_2.\label{rec_L_end}
\end{eqnarray}
These give the initial conditions when one solve the recurrence relation (\ref{rec_L}).
Therefore, any $n$-th degree minimal solutions can be written by $\hB_1$ and $\hB_2$ with $\rho_{\pm}$.%
\footnote{
Apparently, in the recurrence relation (\ref{rec_L}), the two types of the minimal solutions seem to be complexly mixed together. 
But, in fact, the recurrence relation itself can be given for each type,
\begin{eqnarray}
&& \hL_{n,i}^{(1)}=
  \rho_{\pm} \hL_{n-1,i\pm1}^{(1)} - \hL_{n-1,i\pm3}^{(1)} - \hL_{n-2,i}^{(1)},\label{rec_L_1}\\
&&\qquad\qquad {\rm for\ \ } i=-n+3,-n+5\cdots,n-5,n-3, {\ \rm and\ \ } \mp(n-1),\nn\\
&& \hL_{n,i}^{(2)}=
  \rho_{\pm} \hL_{n-1,i\pm1}^{(2)} - \hL_{n-1,i\pm3}^{(2)} - \hL_{n-2,i}^{(2)},\label{rec_L_2}\\
&& \qquad\qquad {\rm for\ \ }i=-n+4,-n+6\cdots,n-6,n-4, {\ \rm and\ \ } \mp(n-2).\nn
\end{eqnarray}
The exceptions  (\ref{1}), (\ref{2}) are 
\begin{eqnarray}
&&\hL_{n-1,\pm n}^{(1)}\equiv \hL_{n,\pm(n-2)}^{(2)},\label{1_1}\\
&&\hL_{n-1,\pm(n-1)}^{(2)}    \equiv \hL_{n-1,\pm(n-2)}^{(1)}.\label{2_2}
\end{eqnarray}
The mixing of the two types occurs at the edges of ranges of $i$ through the exceptions (\ref{1_1}) and (\ref{2_2}).
The lowest degree solution of each type is the fundamental solution, $B_1^{+}$ or $B_2^{+}$.
}

\clearpage
%%%%%%%%%%%%%%%%%%%%%%%%%%%%%%%%%%%%%%%%%%%%%%%%%%%%%%%%%%%%%%%%%%%%
%                                                                                                   %
%                                                                                                   %
%                                                                                                   %
%              Structure of general solution                                                % 
%                                                                                                   %
%                                                                                                   % 
%                                                                                                   %
%%%%%%%%%%%%%%%%%%%%%%%%%%%%%%%%%%%%%%%%%%%%%%%%%%%%%%%%%%%%%%%%%%%%
\vspace{5mm}
\section{General solution of CLR for naive symmetric difference operator}
In this section, we construct the general solution of the CLR for the naive symmetric difference operator. 
In \cite{Kato-2}, the general form of the product $M\Delta$ is given in terms of a two variable holomorphic function, while
we deal with problems concerning the general form of $M$ for the given $\Delta$.

% which is given in the cyclic invariant annulus ${\cal C}_K$.  
%In the previous section and appendix B, any polynomial solution was obtained. Then it was expected that   the general solution can be given by    (A) a linear combination of the two fundamental solutions  with the cyclic invariant holomorphic functions, and (B) a linear combination of the minimal solutions  with complex coefficients.
%In this section, we prove two theorems which assert that these statements are true. 

%\vspace{5mm}
\subsection{Theorem A}
We show the theorem stating that the general solution  of the CLR for the naive symmetric difference operator $\Delta_s$
can be expressed as a linear combination of the fundamental solutions $B_i$ with the cyclic invariant holomorphic functions $f_i$.
Note that some symbols which we use for simplicity of the notation are explained in the footnote.
\footnote{
As defined in  (\ref{M_pm}), 
${\cal M}^{+}(\Delta_s,{\cal C}_K)$ is the solution space of the CLR (\ref{CLR_w_rep})
for the naive symmetric difference operator $\hD_s$ (\ref{D_naive}) with the cyclic annuals ${\cal C}_K$ (\ref{cyclic_annulus}).
 In addition,  
${\cal H}_{\rm cyc}^{+}({\cal C}_K)$, is the space of the cyclic invariant holomorphic functions (\ref{H_pm}). 
$B_i$ are the fundamental solutions (\ref{B1}) and (\ref{B2}).}

%The theorem A is given as follows: 
%
\begin{thm_A}
%Let $M\in{\cal M}^{+}(\Delta_s,{\cal C}_K)$ for a positive constant $K$.
Let $K$ be a positive constant and let $M\in{\cal M}^{+}(\Delta_s,{\cal C}_K)$.
Then, there exist $f_{i} \in {\cal H}_{\rm cyc}^{+}({\cal C}_K)$ such that $\hM$ can be uniquely expressed as 
 \begin{eqnarray}
 \hM = \sum_{i=1}^{2}f_{i}\hB_i,
  \label{GS_exp}
 \end{eqnarray}
where $\hB_i$ are the corresponding fundamental solutions.% of ${\cal M}^{+}(\Delta_s,{\cal C}_K)$.
\end{thm_A}
%

%%%%%%%%%%%%%%%%%%%%%%%%%%%%%%%%%%%%%%%%%%%%%%%%%%%%%%%%%%%%%%%%%%%%
%                                                                                                   %
%                                                                                                   %
%                                                                                                   %
%                         Proof of theorem A                                               % 
%                                                                                                   %
%                                                                                                   % 
%                                                                                                   %
%%%%%%%%%%%%%%%%%%%%%%%%%%%%%%%%%%%%%%%%%%%%%%%%%%%%%%%%%%%%%%%%%%%%

\noindent
{\it Proof:}
Once $M$ is given by (\ref{GS_exp}), $f_{i}$ are unique: 
Suppose $f_i^{\prime} (\ne f_{i})$ are other functions satisfying (\ref{GS_exp}). Then $\sum_{i}(f_{i}-f_{i}^{\prime})B_{i}=0$. 
We find the contradiction $f_{i}^{\prime}=f_{i}$ using the two properties of the delta operators, (\ref{delta_lin}) and (\ref{delta_B}). 
Consequently, all we need to do is show that $\hM$ can be expressed as (\ref{GS_exp}). 

Now, let us choose $f_{i} = \delta_{i} M \in{\cal H}_{\rm cyc}^{+} ({\cal C}_K)$.
Then,  
\begin{eqnarray}
&&\hspace{-1.5cm} \sum_{i=1}^2 f_{i}(w,z) B_{i}(w,z)\nn\\
 &=&
 \frac{\left(\sum_{i,j}\epsilon_{ij}B_{i}(w,z)B_{j}\left(w,\frac{1}{wz}\right)\right)M\left(z,\frac{1}{wz}\right)-(w\leftrightarrow z)}
      {2\iota(w,z)\Delta_s\left(\frac{1}{wz}\right)},
      \label{M_f}\\\nn\\
 &=& -\frac{\Delta_s\left(w\right)M\left(z,\frac{1}{wz}\right)
             +\Delta_s\left(z\right)M\left(w,\frac{1}{wz}\right)}
             {\Delta_s\left(\frac{1}{wz}\right)}, 
             \label{M_BiBj}  \\\nn\\
 &=& M(w,z)-
    \frac{\left(\Delta_s\left(\frac{1}{wz}\right)M(w,z)
    +\Delta_s\left(w\right)M\left(z,\frac{1}{wz}\right)
             +\Delta_s\left(z\right)M\left(w,\frac{1}{wz}\right)\right)}
             {\Delta_s\left(\frac{1}{wz}\right)}, 
             \label{M_BiBj} \\\nn\\
 &=& M(w,z) \label{M_{CLR}}.
\end{eqnarray}
We used   the definition of $f_{i}$ (\ref{f_delta}), the identity (\ref{cyclic_Euler}) and the CLR (\ref{CLR_w_rep}). 
Therefore the theorem holds.  %\qed
 $\qquad{\mathbf \square}$
\vspace{5mm}

%The theorem and the corollary tell us that  
%the number of the fundamental solutions in each sector of ${\cal M}^{\pm}(\Delta_s,{\cal C}_K)$ 
%is two.   In appendix C, it is found that the number for any difference operator is also two.

 Conversely, for given $f_i \in {\cal H}_{\rm cyc}^{+}({\cal C}_K) $, we can find  a local solution of the CLR using the theorem.%
 \footnote{ If the normalized solutions are needed, an extra constraint has to be imposed on $f_i$, $6\sum_{i=1}^{2}f_i(3,3)=1$.}
As a consequence of the theorem,
 ${\cal M}^{+}(\Delta_s,{\cal C}_K)$ and ${\cal H}_{\rm cyc}^{+}({\cal C}_K) \times {\cal H}_{\rm cyc}^{+}({\cal C}_K)$ are in a one-to-one correspondence.
\footnote{
The general elements  of ${\cal M}^{-}(\Delta_s,{\cal C}_K)$ (\ref{M_pm}) can be expressed as (\ref{GS_exp}) with 
\begin{eqnarray}
 \hB_{1}(w,z) = w-\frac{1}{w}-z+\frac{1}{z},\qquad
 \hB_{2}(w,z) = \frac{w}{z}-\frac{z}{w}.
\end{eqnarray}
%The proof is the same with that of  (\ref{GS_exp}).
} 
 %

%

%%%%%%%%%%%%%%%%%%%%%%%%%%%%%%%%%%%%%%%%%%%%%%%%%%%%%%%%%%%%%%%%%%%%
%                                                                                                   %
%                                                                                                   %
%                                                                                                   %
%              Structure of general solution : theorem B                                 % 
%                                                                                                   %
%                                                                                                   % 
%                                                                                                   %
%%%%%%%%%%%%%%%%%%%%%%%%%%%%%%%%%%%%%%%%%%%%%%%%%%%%%%%%%%%%%%%%%%%%
\vspace{5mm}
\subsection{Theorem B}
In this section, we show that the general solution can be expressed as in terms of the minimal solutions $L_{ni}$, (\ref{min_1}) and (\ref{min_2}), with exponentially decaying coefficients.   
Conversely, once such coefficients 
are given,  we can obtain a local solution of the CLR by using the theorem.%
\footnote{
Here, the normalization condition (\ref{cond_hM_2}) is not assumed.
We need an extra  constraint on $\alpha_{ni}$, $6\sum_{n\in{\mathbb N_+}}\sum_{i=-n+1}^{n-1}\alpha_{ni}=1$, for the normalized solutions.}

\begin{thm_B}
Let $K$ be a positive constant and let $M\in{\cal M}^{+}(\Delta_s,{\cal C}_K)$.
Then,  $\hM$ can be uniquely expressed as 
 \begin{eqnarray}
  \hM = 
  \sum_{n=1}^{\infty}
  \sum_{i=-n+1}^{n-1} \alpha_{ni} \hL_{ni},
  \label{GS_min_exp}
 \end{eqnarray}
 where $\hL_{ni}$ are the minimal solutions of degree $n$, 
and  $\alpha_{ni}$ are complex constants.
 The right-hand side of (\ref{GS_min_exp}) absolutely and uniformly converges on every compact subset of ${\cal C}_K$, and 
  $\alpha_{ni}$ satisfy 
\begin{eqnarray}
\left|\alpha_{ni}\right|
 \leq
C^{\prime}\exp\left(-K^{\prime}n \right),
\label{alpha_cond}
\end{eqnarray}
 for $0<{}^{\forall} K^{\prime}<K$, 
 where  $C^{\prime}$ is a  positive constant.
 \end{thm_B}
%
%

%%%%%%%%%%%%%%%%%%%%%%%%%%%%%%%%%%%%%%%%%%%%%%%%%%%%%%%%%%%%%%%%%%%%
%                                                                                                   %
%                                                                                                   %
%                                                                                                   %
%                         Proof of uniqueness                                                % 
%                                                                                                   %
%                                                                                                   % 
%                                                                                                   %
%%%%%%%%%%%%%%%%%%%%%%%%%%%%%%%%%%%%%%%%%%%%%%%%%%%%%%%%%%%%%%%%%%%%

\noindent
{\it Proof:}
First, once  $M$ is written as the right-hand side of (\ref{GS_min_exp}), we can show that the expression is unique from the following 
lemma,
\begin{eqnarray}
 \sum_{n=1}^{\infty}
 \sum_{i=-n+1}^{n-1} \alpha_{ni} \hL_{ni} =0
 \quad&\Longleftrightarrow&\quad
 \alpha_{ni} =0,
 \label{uniq_A}
\end{eqnarray}
for any complex constants $\alpha_{ni}$ satisfying (\ref{alpha_cond}). 
If there exist other coefficients $\alpha_{ni}^{\prime} (\neq \alpha_{ni})$ representing $\hM$, then it follows that $\sum_{n,i}(\alpha_{ni}-\alpha_{ni}^{\prime})\hL_{ni}=0$.  
If the lemma holds,
 it is immediately apparent that $\alpha_{ni}^{\prime}=\alpha_{ni}$. Therefore, once $M$ is given by (\ref{GS_min_exp}), the expression is unique if the lemma is true.

Let us prove the lemma.  Clearly, $\sum_{n,i} \alpha_{ni}\hL_{ni}=0$ if $\alpha_{ni}=0$. 
However, the converse is non-trivial and 
 can be shown inductively.  
Note that $\sum_{n,i} \alpha_{ni}\hL_{ni}=0$ means that each monomial of $w, z$ in $\sum_{n,i} \alpha_{ni}\hL_{ni}$ should vanish. 
Because only $\hL_{10}$ has the  zero-th degree monomial (a constant),   so that $\alpha_{10}=0$. 
Now assume that $\alpha_{mi}=0$, $m\leq n$. 
 Then, the $n$-th degree monomials should satisfy  
\begin{eqnarray}
 \sum_{i=0}^{n} \alpha_{n+1,n-2i}\frac{w^i}{z^{n-i}}
+\sum_{i=1}^{n} \alpha_{n+1,n+1-2i}\left(w^nz^{n+1-i}+\frac{1}{w^nz^{i}}\right)+(w\leftrightarrow z)=0. 
\label{uniq_cond_N}
\end{eqnarray}
All of the monomials in  (\ref{uniq_cond_N}) are independent of each other, so that $\alpha_{n+1,i}=0$, $|i|\leq n$. Therefore, $\alpha_{ni}=0$ for all $n\in{\mathbb N_+}$ and the lemma holds.

In the rest of this section, we show that $M$ can be expressed as (\ref{GS_min_exp}). It follows from theorem A that
\begin{eqnarray}
 \hM = \sum_{i=1}^{2} f_{i} B_{i},
\end{eqnarray}
where $f_{i}$ are the elements of ${\cal H}_{\rm cyc}^{+}({\cal C}_K)$. From  (\ref{cyclic_Laurent_plus}), $f_i$ can be expanded as 
\begin{eqnarray}
 f_{i} = c_{i}+\sum_{n=1}^{\infty} \sum_{k=0}^{n} c_{ink}p_{nk}^{+},
 \label{f_expand}
\end{eqnarray} 
and, for  $0<{}^{\forall}K^{\prime}<K$,
\begin{eqnarray}
 |c_{ink}| \le C e^{-K^{\prime}n},
\end{eqnarray} 
where $C$ is a positive constant, as well as (\ref{cond_local_c_ni}). 
The right-hand side of (\ref{f_expand})  is absolutely and uniformly convergent on every compact subset of  ${\cal C}_K$. 

Using the relations, 
\begin{eqnarray}
 p_{nk}^{+} B_{1} &=& \frac{1}{N_{nk}}(L_{n,n-2k-1}+L_{n,n-2k}+L_{n,n-2k+1}\nn\\
              & & \qquad \quad +L_{n+1,n-2k-1}+L_{n+1,n-2k}+L_{n+1,n-2k+1}),\\
 p_{nk}^{+} B_{2} &=& \frac{1}{N_{nk}}(L_{n-1,n-2k}+L_{n,n-2k-2}+L_{n,n-2k+2}\nn\\
              & & \qquad \quad +L_{n+1,n-2k-2}+L_{n+1,n-2k+2}+L_{n+2,n-2k}),
\end{eqnarray} 
%
%for $n\geq 0$, $k=0,1,\cdots,n$, 
where $N_{nn}=N_{n0}=2$ and the other $N_{nk}$ are unity, and 
\begin{eqnarray}
 L_{n,\pm n}=L_{n,\pm(n-1)},\quad L_{n,\pm(n+1)}=L_{n+1,\pm(n-1)},\quad L_{n,\pm(n+2)}=L_{n+1,\pm(n-2)},
\end{eqnarray} 
$\hM$ is easily shown to be 
\begin{eqnarray}
&& \hM = \sum_{n=1}^{\infty} \sum_{i=-n+1}^{n-1} \alpha_{ni} L_{ni},
\end{eqnarray}
where each $\alpha_{ni}$ is the sum of  six $c_{imk}\ (n-2\leq m \leq n+1)$ with the same or different signs
 and  satisfy 
\begin{eqnarray}
 |\alpha_{ni}| \le C^{\prime} e^{-K^{\prime} n},
\end{eqnarray}
where $C^{\prime}$ is a positive constant. \qquad $\Box$

%%%%%%%%%%%%%%%%%%%%%%%%%%%%%%%%%%%%%%%%%%%%%%%%%%%%%%%%%%%%%%%%%%%%
%                                                                                                   %
%                                                                                                   %
%                                                                                                   %
%                         Summary                                                              % 
%                                                                                                   %
%                                                                                                   % 
%                                                                                                   %
%%%%%%%%%%%%%%%%%%%%%%%%%%%%%%%%%%%%%%%%%%%%%%%%%%%%%%%%%%%%%%%%%%%%
\vspace{5mm}
\section{Summary}

It is well known  that the realization of 
full supersymmetry on the lattice is difficult  due to the violation of the Leibniz rule for the local difference operators. 
The cyclic Leibniz rule creates new possibilities to realize the lattice supersymmetry beyond the violation. 
In this paper, we have studied the structure of the general solution of the 2-body CLR for the naive symmetric difference operator, and shown that   the general solution can be uniquely expressed as (A)  a linear combination of the two fundamental solutions with cyclic invariant coefficients, and (B) a linear combination of the minimal solutions with complex coefficients.

We have prepared several mathematical definitions and tools to obtain the results. Then, it was  important to stand a point of view respecting the cyclic symmetry. As the manifestation of the symmetry, almost everything we introduced has $cyclic$ in its names: cyclic locality, cyclic invariant annulus and so on.
Moreover, we have used the theory of functions of several complex variables 
everywhere in the proofs. 
This implies that the CLR has a solid mathematical background. Therefore, it is reasonable to regard the CLR as a systematic and constructive  approach to the lattice supersymmetry.

Toward the constructive theory of the lattice supersymmetry based on the CLR, there still remains much to be done. 
In particular, extensions to higher dimensional theory and gauge theory must be the most important problems to be solved.  
Even in one dimension, how to construct the general solution for any difference operator remains as an open question. 
In appendix C, at the beginning of the extension,
we showed that the number of fundamental solutions of the 2-body CLR does not depend on the type of the difference operators and is always two. 
Even in the higher dimensional theory and the gauge theory, if some relations are cyclic invariant or covariant, several concepts introduced in this paper will be useful.

%
% Acknowledgment
%
\vspace{5mm}
\section*{Acknowledgment}
We would like to thank M. Kato, M. Sakamoto and H. So for their discovery of the cyclic Leibniz rule. 
This work is supported in part by the Grant-in-Aid for Scientific Research (No.24740143(N.U.)) by the Japanese  Ministry of Education, Science, Sports and Culture.

%
% Appnendix
%
\appendix

%%%%%%%%%%%%%%%%%%%%%%%%%%%%%%%%%%%%%%%%%%%%%%%%%%%%%%%%%%%%%%%%%%%%
%                                                                                                   %
%                                                                                                   %
%                                                                                                   %
%                       Appendix. .A                                                            % 
%                                                                                                   %
%                                                                                                   % 
%                                                                                                   %
%%%%%%%%%%%%%%%%%%%%%%%%%%%%%%%%%%%%%%%%%%%%%%%%%%%%%%%%%%%%%%%%%%%%
\vspace{5mm}
\section{Polynomial solution of degree six}
In this appendix, we construct the general 6-th degree polynomial solution of the CLR for the naive symmetric difference operator  
by solving the recurrence relation (\ref{Amn}). 
The construction can be easily extended to any degree. However, it is purely  algebraic and unsuitable for understanding the structure of the solution. 
Instead,  the table representation of a solution given 
in appendix B is useful for discussing the structure.

The number of $S_{kl}$ associated with a polynomial solution of degree $n$ is $\frac{1}{2}(3n+2)(n+1)$, while the number of constraints from the recurrence relation is $\frac{1}{2}(n+2)(n+3)$. As discussed in the footnote of section 3.2, at least two of the constraints  are linearly dependent to the others. Thus, at most $n^2$ initial values are needed to solve the recurrence relation if there are no further linearly dependent relations.

For 6-th degree solutions, 
first, let us prepare complex constants $\alpha_{ni}$ $(n\geq1, |i|\leq n-1)$ satisfying
\begin{eqnarray}
 &&\alpha_{n,i}=0, \qquad{\rm \  for\ \ } n> 6,
\end{eqnarray}
and set the 36 initial values,
\begin{eqnarray}
 &&S_{ni} = \alpha_{n,2i-n+1}+\alpha_{n+1,2i-n-1}, \qquad  (i=1,\cdots,n-1),
 \label{ini_val_s}\\
 &&S_{n0} =  \alpha_{n,-n+1}+\alpha_{n+1,-n},\\
 &&S_{i,-(n-i)} = \alpha_{n,n-2i}+\alpha_{n+1,n-2i}, \qquad (i=1,\cdots,n-1), 
\end{eqnarray}
for $n=1,\cdots,6$.

One can explicitly solve $A_{ni}=0$ in the order of $n=7,6,\cdots,2$ and $A_{10}=0$ 
and obtain all of $S_{kl}$ as follows:
\begin{eqnarray}
  &&S_{-n,-i} = \alpha_{n,-2i+n-1}+\alpha_{n+1,-2i+n+1}, \qquad  (i=1,\cdots,n-1),\\
 &&S_{-n,0} =  \alpha_{n,n-1}+\alpha_{n+1,n},
\end{eqnarray} 
and 
\begin{eqnarray}
 &&S_{n,n} = 2\alpha_{n+1,n-1},\\
 &&S_{-n,-n} = 2\alpha_{n+1,-n+1},\\
 && S_{00} = 2\alpha_{10},
 \label{ini_val_e}
\end{eqnarray}
for $n=1,\cdots,6$. 
The other two relations $A_{11}=A_{00}=0$ consistently hold for (\ref{ini_val_s})-(\ref{ini_val_e}). 

As a result, the polynomial solution of degree six is given by
\begin{eqnarray}
 \hM_6 = \sum_{n=1}^{6}\sum_{i=-n+1}^{n-1}\alpha_{ni} \hL_{ni},
\end{eqnarray}
where $\hL_{ni}$ are given in (\ref{min_1}), (\ref{min_2}). 
Moreover, it can be also  expressed as
\begin{eqnarray}
 \hM_6 = \sum_{i=1}^{2}f_{i}(\rho_+,\rho_-)\hB_i,
\end{eqnarray} 
where $f_{i}\ (i=1,2)$ are the $(6-i)$-th degree polynomials of $\rho_{\pm}$ (\ref{rho+})  and  (\ref{rho-}).
From the recurrence relation of $L_{ni}$ (\ref{rec_L}), $f_{i}$ are uniquely determined from $\alpha_{ni}$.

%%%%%%%%%%%%%%%%%%%%%%%%%%%%%%%%%%%%%%%%%%%%%%%%%%%%%%%%%%%%%%%%%%%%
%                                                                                                   %
%                                                                                                   %
%                                                                                                   %
%                       Appendix. B                                                            % 
%                                                                                                   %
%                                                                                                   % 
%                                                                                                   %
%%%%%%%%%%%%%%%%%%%%%%%%%%%%%%%%%%%%%%%%%%%%%%%%%%%%%%%%%%%%%%%%%%%%
\vspace{5mm}
\section{Finite dimensional versions of the theorems}
The finite degree versions  of theorems A and B, which correspond to the ultra local solutions, can be derived from the theorems themselves. 
However, a table representation of the solution
gives us a somewhat depth understanding of its structure.
In this appendix, we prove the finite versions by using another approach based on the table representation.

%%%%%%%%%%%%%%%%%%%%%%%%%%%%%%%%%%%%%%%%%%%%%%%%%%%%%%%%%%%%%%%%%%%%
%                                                                                                   %
%                                                                                                   %
%                                                                                                   %
%                     Representaion using tables                                           % 
%                                                                                                   %
%                                                                                                   % 
%                                                                                                   %
%%%%%%%%%%%%%%%%%%%%%%%%%%%%%%%%%%%%%%%%%%%%%%%%%%%%%%%%%%%%%%%%%%%%

% figs. of B_1 and B_2 
\begin{figure}[t!]
\begin{center}
 {\includegraphics[width=0.4\columnwidth,clip]{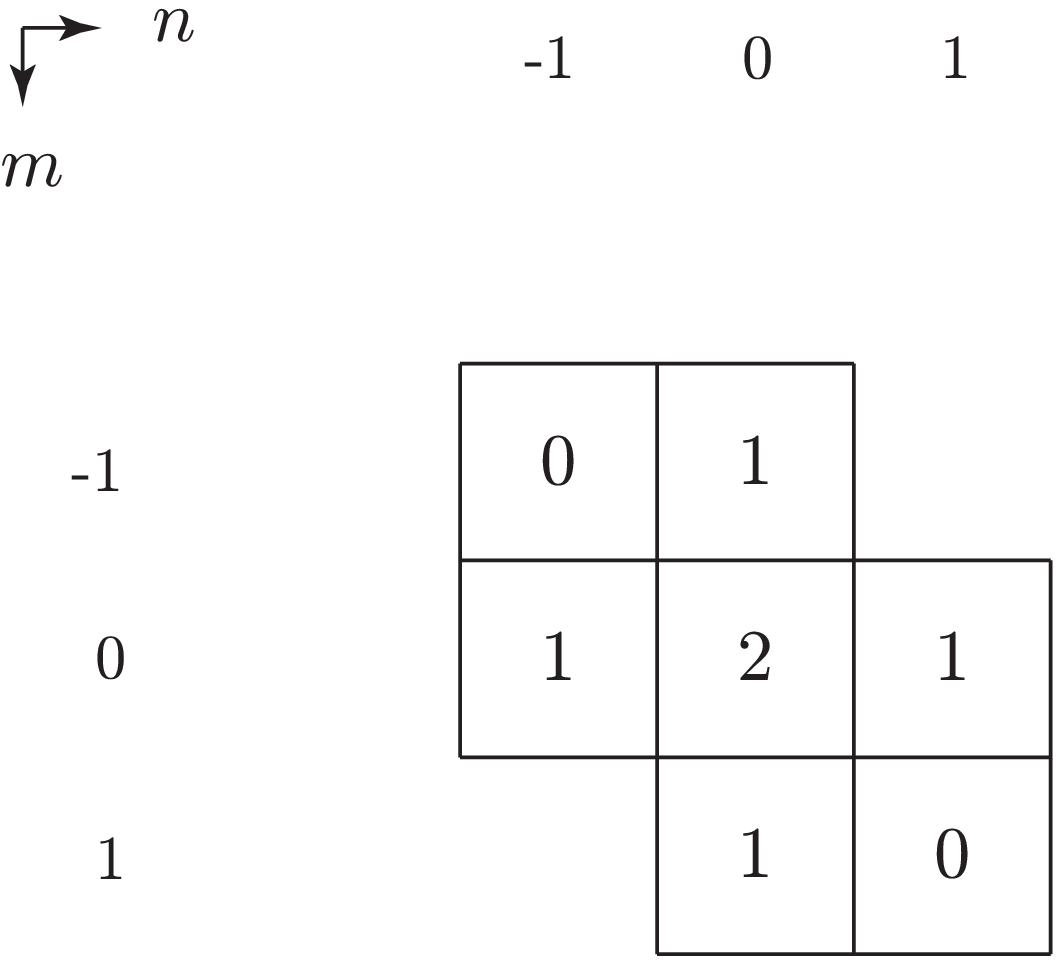}}
  \hspace{20mm}
 {\includegraphics[width=0.4\columnwidth,clip]{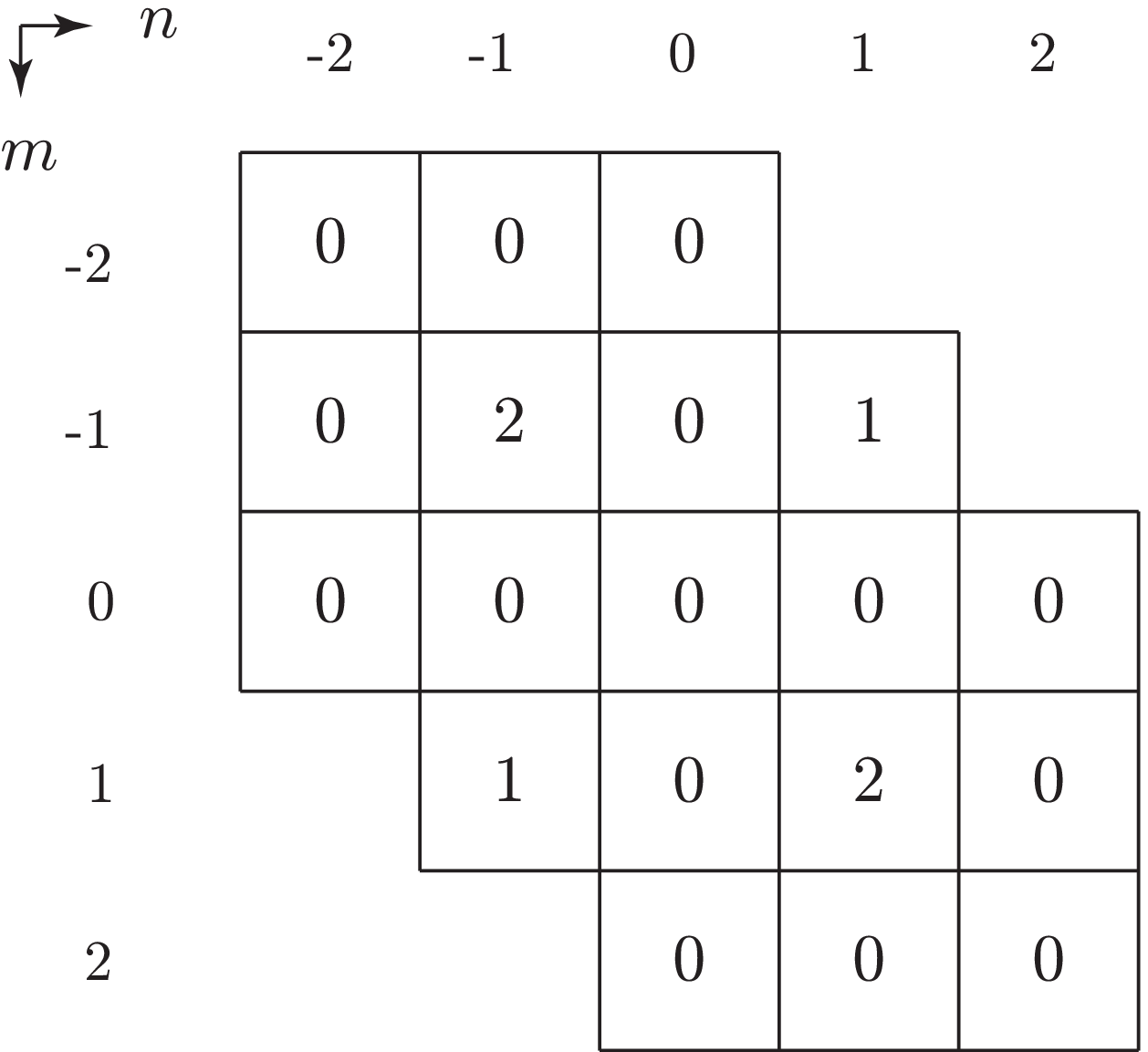}}
 \caption{Table representations of $\hB_1$ (left) and $\hB_2$ (right)
  \label{fig_B12}}
\end{center}
\end{figure}

\subsection{Table representation of the solutions}
% figs. of hM_2 
\begin{figure}[t!]
\begin{center}
 {\includegraphics[width=0.4\columnwidth,clip]{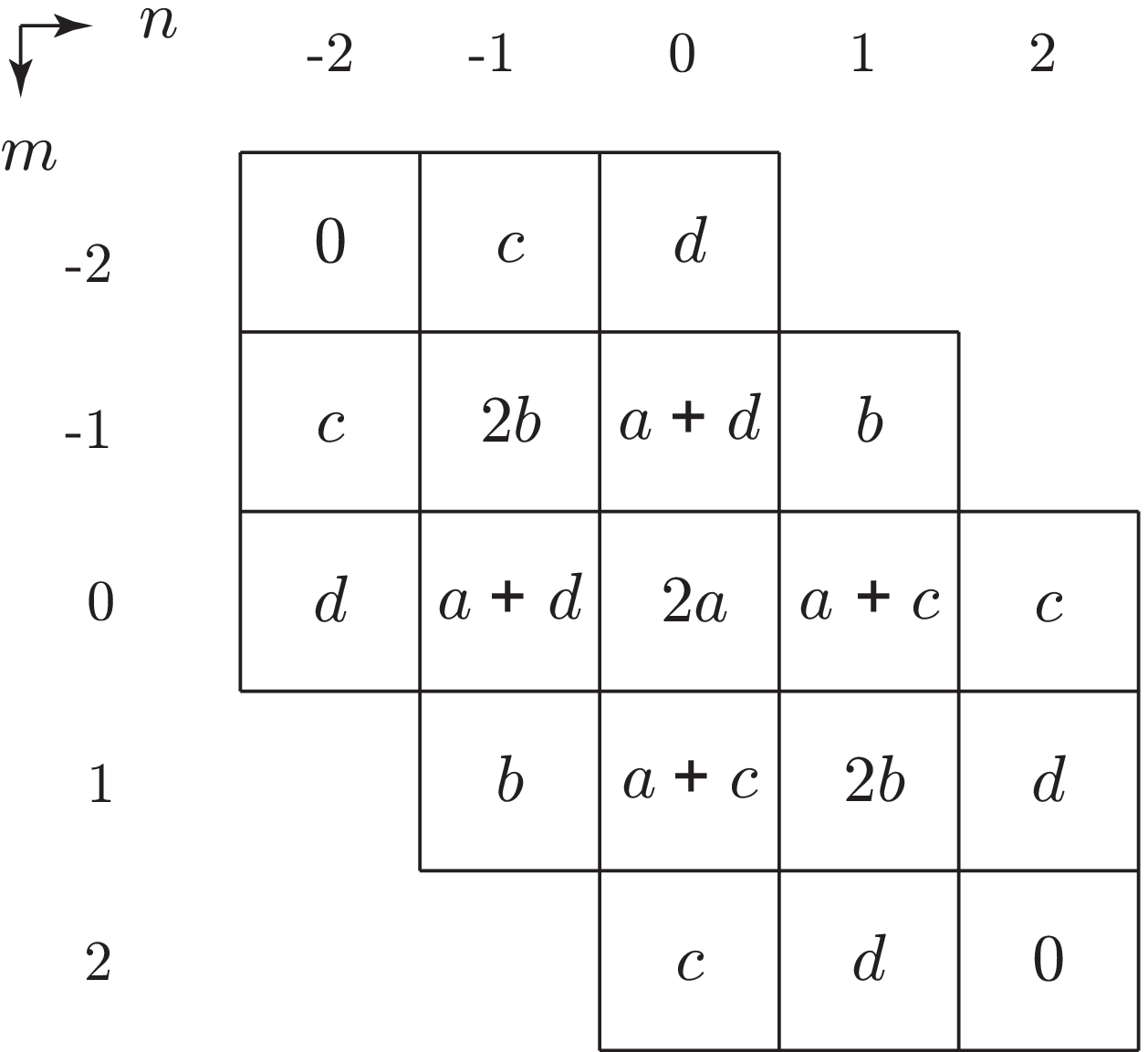}}
\caption{Table representation of $\hM_2$\label{fig_hM_2}}
\end{center}
\end{figure}
%

% fig. of hM_N 
\begin{figure}[t!]
\begin{center}
 {\includegraphics[width=0.3\columnwidth,clip]{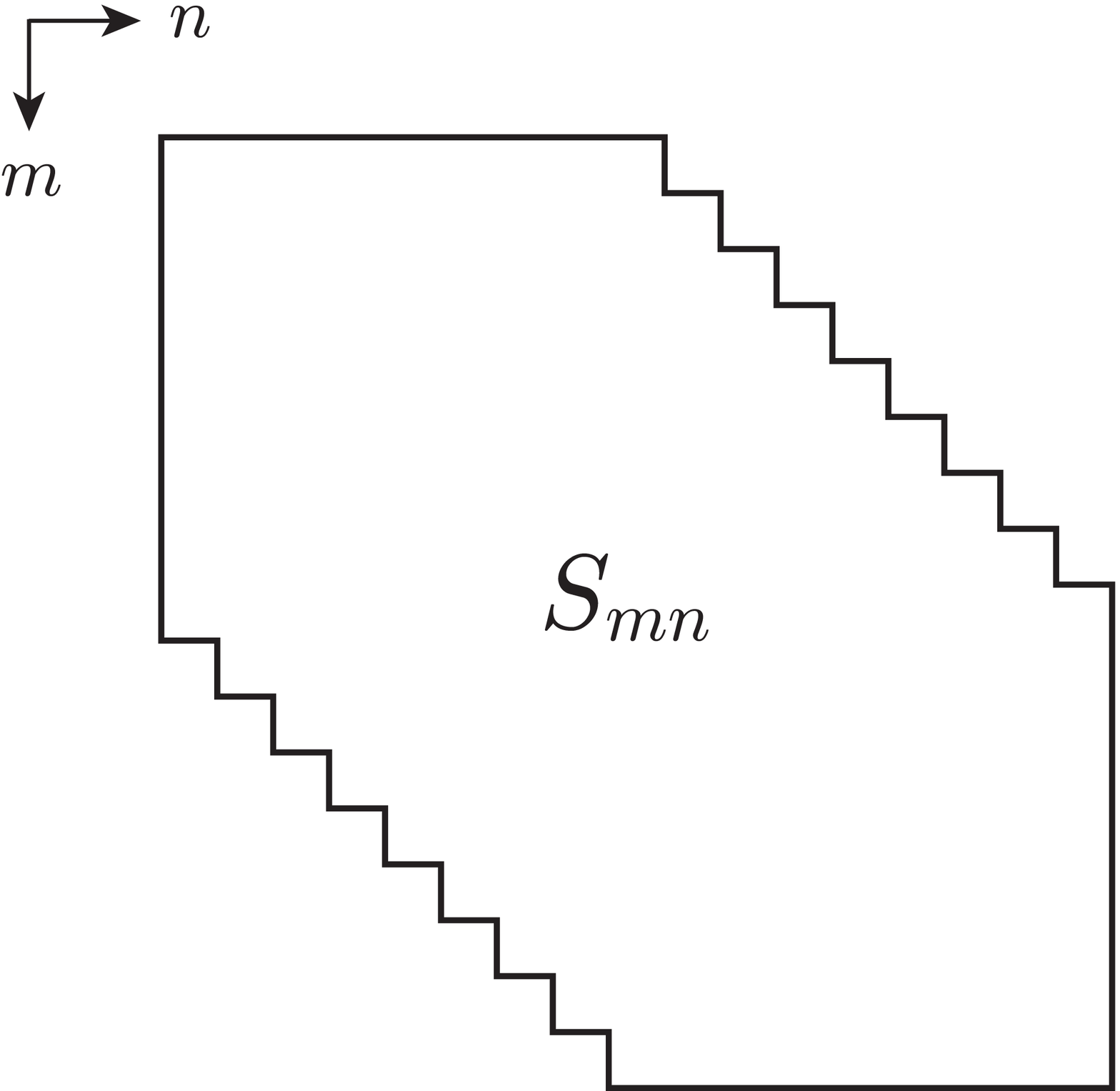}}
 \hspace{20mm}
 {\includegraphics[width=0.3\columnwidth,clip]{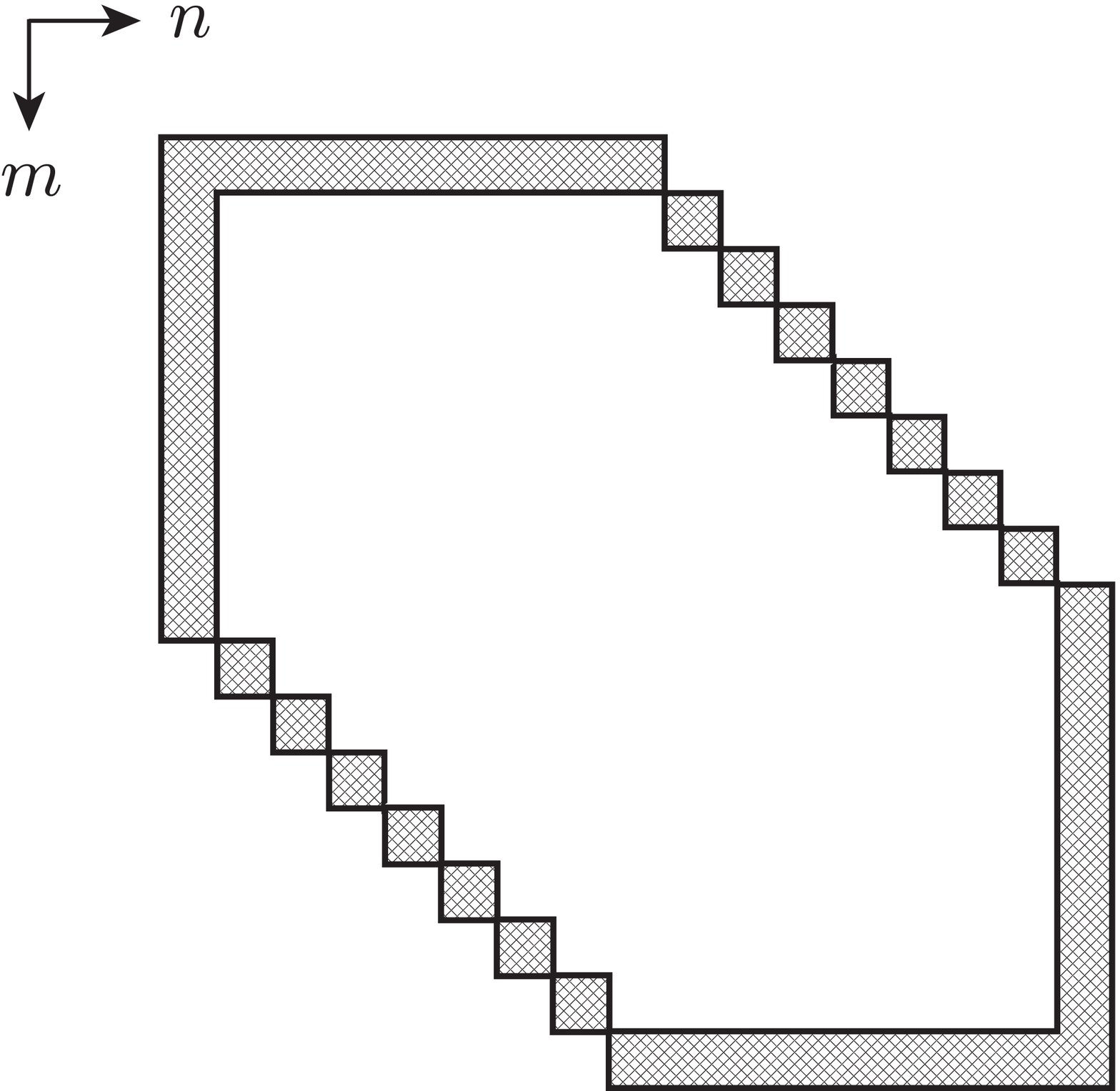}}
\caption{Shape of $\hM_n$ (left) and $n$-th boundary $R_n$ (right)\label{fig_hM_N}}
\end{center}
\end{figure}

We introduce the table representation of the solutions, which gives us a clear insight into the proof of the lemma given in the next subsection.

%Let us draw $S_{kl}$ as a table.
The two fundamental solutions, $\hB_1$ (\ref{B1}) and $\hB_2$ (\ref{B2}), are represented in figure  \ref{fig_B12}. Each value in the cell $(k,l)$ corresponds to the coefficient $S_{kl}$ and the other values outside the tables  are zero. Figure \ref{fig_hM_2} shows the general 2nd degree polynomial solution, $\hM_2$ (\ref{hM_2}). 
Similarly, figure \ref{fig_hM_N} represents the general solution of degree $n$. 
The shape of the table is not a square but a hexagon due to the definition of the cyclic invariant degree, (\ref{order}).
The shaded region denotes the set of the $n$-th degree monomials,
% (see figure \ref{fig_R_N_def}),
%
\begin{eqnarray}
 R_{n} = \big\{ \, (k,l) \, | {\rm \ degree\ of\ }w^kz^l = n, \, {\rm for} \ k,l\in {\mathbb Z} \, \big\},
\end{eqnarray}
which will be referred to as $n$-th boundary.
%
%The cells on and inside the $n$-th boundary are determined by the recurrence relation, $A_{mi}=0$ for $m\leq n+1$,  the ones outside the $n$-th boundary are zero by the definition. 

% fig. of \rho_+ and \rho_- 
\begin{figure}[t!]
\begin{center}
%\subfigure[]
 {\includegraphics[width=0.25\columnwidth]{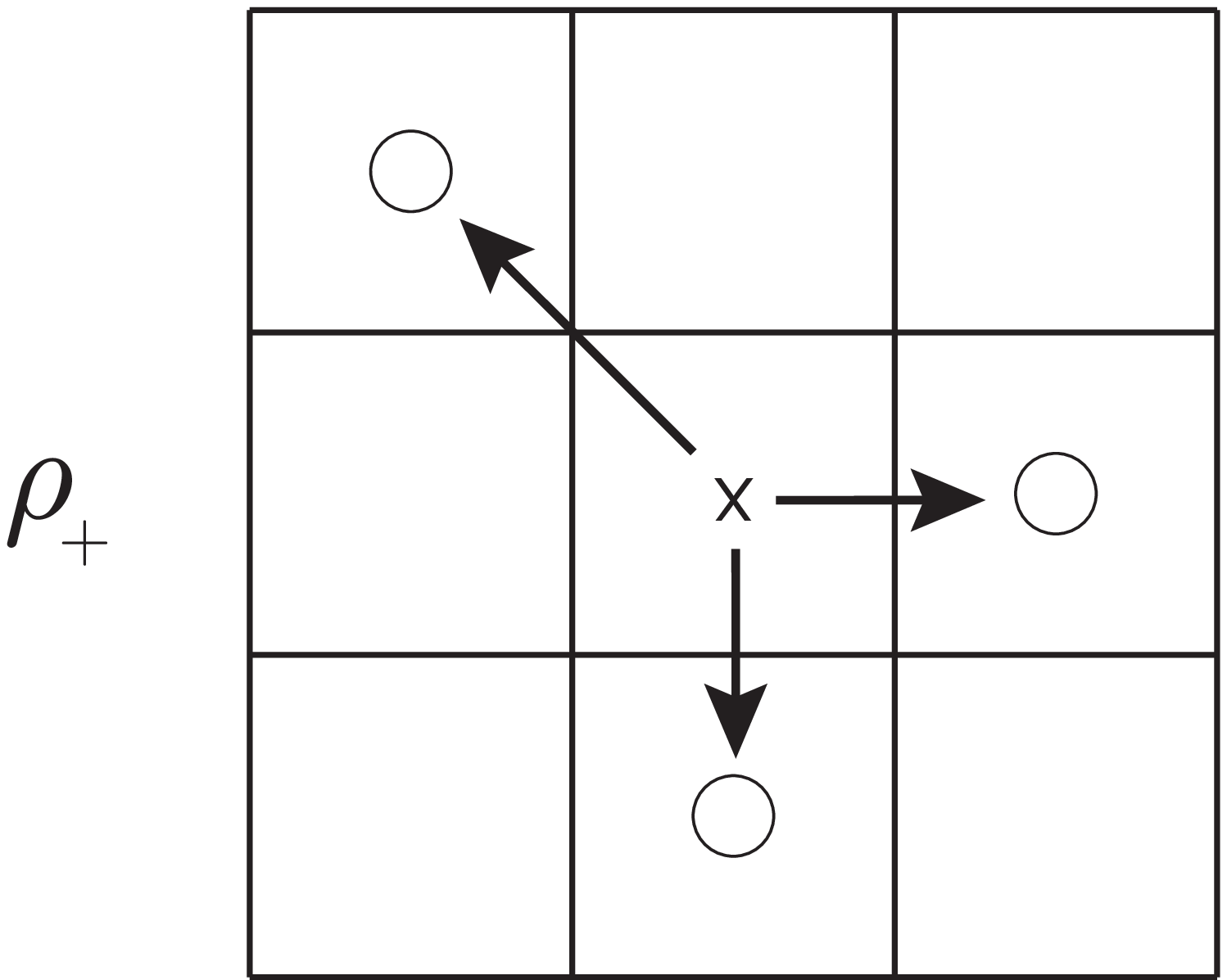}\label{fig_rho_+}}
\hspace{20mm}
%\subfigure[]
 {\includegraphics[width=0.25\columnwidth]{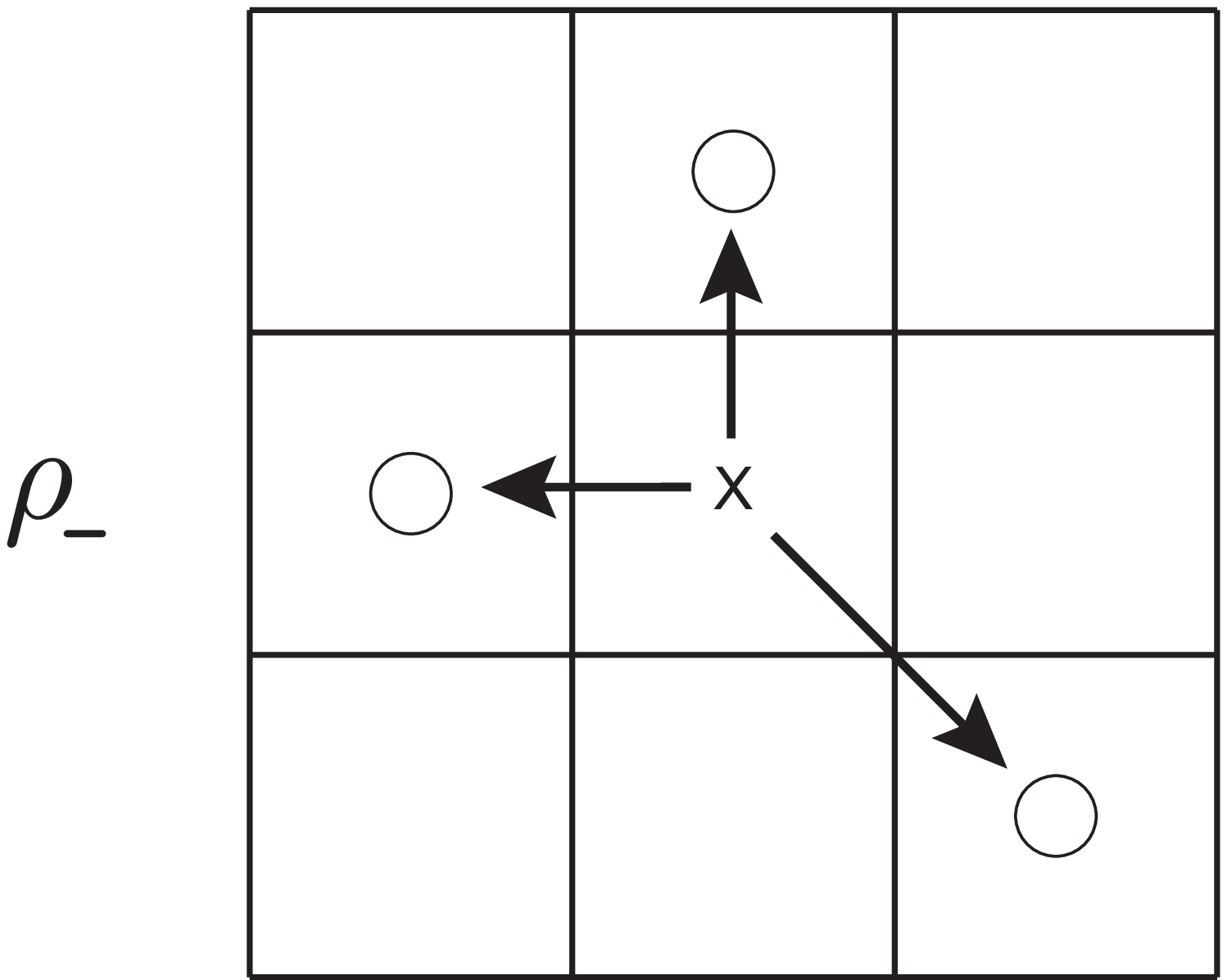}\label{fig_rho_-}}
\caption{Geometrical meaning of $\rho_+$ (left) and $\rho_-$ (right).  $\rho_{\pm}$ make the coefficient $S_{kl}$ in the cell $(k,l)$ move to the three cells $(k\pm1,l), (k,l\pm1), (k\mp1,l\mp1)$.\label{fig_g}}
\end{center}
\end{figure}
%

%Next, let us consider  geometrical meanings of the operations of $\rho_{\pm}$.
Any $n$-th degree monomial multiplied by $\rho_{\pm}$ becomes the $n+1$-th degree polynomials that consist of the three monomials. In figure \ref{fig_g}, 
the cross denotes the original $n$-th degree monomial, while the three circles denote the three monomials after the multiplications of $\rho_{\pm}$.
For the $n$-th degree solution $\hM_{n}$, $\rho_{\pm}$ make the value in each cell $(k,l)$ move to the three cells $(k\pm1,l), (k,l\pm1), (k\mp1,l\mp1)$. The new $n+1$-th solutions $\tilde M^\pm_{n+1}\equiv \rho_{\pm}\hM_{n}$ are given by the superpositions of the transferred values,
\begin{eqnarray}
 \rho_{\pm} &:& \ S_{k,l}\ \longmapsto \ \tilde S^{\pm}_{k,l}=S_{k\mp1,l}+S_{k,l\mp1}+S_{k\pm1, l\pm1},
\end{eqnarray}
where $S_{kl}$ and $\tilde S_{kl}^{\pm}$ are matrices relevant to $\hM_{n}$ and $\tilde M_{n+1}$, respectively.
By using these $\rho_\pm$-moves, $\hM_2$ (\ref{hM_2_gen}) can be easily shown to be figure \ref{fig_hM_2}.

%%%%%%%%%%%%%%%%%%%%%%%%%%%%%%%%%%%%%%%%%%%%%%%%%%%%%%%%%%%%%%%%%%%%
%                                                                                                   %
%                                                                                                   %
%                                                                                                   %
%                       Ultra local solutions                                                  % 
%                                                                                                   %
%                                                                                                   % 
%                                                                                                   %
%%%%%%%%%%%%%%%%%%%%%%%%%%%%%%%%%%%%%%%%%%%%%%%%%%%%%%%%%%%%%%%%%%%%

\vspace{5mm}
\subsection{Lemma A} 
In this appendix, the finite version of theorem A is given without a proof. 
One can easily show it from lemma B given in appendix B.3 and the recurrence relation (\ref{rec_L})

\begin{lem_A}
 Let $\hM_{n}$ be any $n$-th degree polynomial solution of the CLR  for the naive symmetric difference operator, which satisfies $\hM_n(w,z)=\hM_n(z,w)$. Then, 
 it can be  uniquely expressed as,  
 \begin{eqnarray}
  \hM_{n} = \sum_{i=1}^{2}f_{ni}(\rho_+,\rho_-)\hB_i,
  \label{GS}
 \end{eqnarray}
where $\hB_i$ are the two fundamental solutions, and  $f_{ni}$ are  polynomials  of degree $n-i$ in $\rho_{\pm}$.%
\footnote{
The polynomial of $\rho_{\pm}$ in the lemma is the usual one.  
That is, it consists 
 of  the monomials with non-negative power, $\rho_{+}^k\rho_{-}^l\ (k,l\in{\mathbb N})$, and the corresponding degree is $k+l$.
}
\end{lem_A}
%
%There is a difference between the lemma and the theorem : 
% $\hM_{n}$  has a finite degree $n$, so that  $f_{ni}$ are polynomials.   
%The polynomial of $\rho_{\pm}$ in the lemma is the usual one.  
%That is, it does consist 
% of  monomials with non-negative power, $\rho_{+}^k\rho_{-}^l\ (k,l\in{\mathbb N})$, and the corresponding degree is defined as a maximal value of $k+l$.

%%%%%%%%%%%%%%%%%%%%%%%%%%%%%%%%%%%%%%%%%%%%%%%%%%%%%%%%%%%%%%%%%%%%
%                                                                                                   %
%                                                                                                   %
%                                                                                                   %
%                       Ultra local solutions                                                  % 
%                                                                                                   %
%                                                                                                   % 
%                                                                                                   %
%%%%%%%%%%%%%%%%%%%%%%%%%%%%%%%%%%%%%%%%%%%%%%%%%%%%%%%%%%%%%%%%%%%%
\vspace{5mm}

\subsection{Lemma B}
We prove the finite version of theorem B.  The table representation of a solution plays a key role in the proof. 
\begin{lem_B}
Let $\hM_{n}(w,z)$ be any $n$-th degree solution of the CLR  for the naive symmetric difference operator, which satisfies $\hM_n(w,z)=\hM_n(z,w)$. 
Then, it has a unique  representation,  
 \begin{eqnarray}
   \hM_n =  
  \sum_{m=1}^{n}
  \sum_{i=-m+1}^{m-1} \alpha_{mi} \hL_{mi},
  \label{GS_min}
 \end{eqnarray}
where $L_{mi}$ are the minimal solutions of degree $m$, 
and $\alpha_{mi}$ are complex constants.%
\footnote{
A  crucial  difference from the infinite version is that 
  $\alpha_{mi}$ $(m\leq n)$ are any constants without (\ref{alpha_cond}).
  This is because $\hM_{n}$  is the ultra local solution.
}
\end{lem_B}
%
%
%
%A  crucial  difference between the finite and the infinite versions is that 
%  $\alpha_{mi}$ $(m\leq n)$ are any constants without (\ref{alpha_cond}),
%  because $\hM_{n}$  is  a ultra local solution.

\noindent
{\it Proof:} We prove the lemma by mathematical induction for $n$. 
Since the two fundamental solutions are polynomials of degree one and two, $n=1, 2$ are the base cases and $n\geq3$ are the inductive steps.

\vspace{5mm}
\noindent
(i) The general solution of degree one was already given in  (\ref{hM_1}).  Using $\hL_{10}$, we find  
\begin{eqnarray}
 \hM_1 = a \hL_{10}, %\quad a\in{\mathbb C}.
\end{eqnarray}
where $a$ is a complex constant.
By setting $\alpha_{10}\equiv a$, one can rewrite $\hM_1$ as (\ref{GS_min}), so that the lemma holds for $n=1$.

For $n=2$,  the general solution  (\ref{hM_2}) can be expressed as  
\begin{eqnarray}
\hM_{2} = a \hL_{10}
        + b \hL_{20}
        + c \hL_{21}
        + d \hL_{2-1},        
\end{eqnarray}
where 
$a, b, c$ and  $d$ are some complex constants.
Then it is possible to show that the lemma holds for $n=2$, by choosing 
$\alpha_{10}\equiv a$, $\alpha_{20}\equiv b$, $\alpha_{21}\equiv c$ and $\alpha_{2-1}\equiv d$.

\vspace{5mm}
\noindent
(ii) 
Let us assume that  the lemma has been established for $n-1$ $(n\geq 3)$. 

First of all, we consider  
the boundary structure of $\hM_{n}$.
In the recurrence relation, the $n$-th boundary cells $R_n$ are included in  $A_{mi}=0$ for $n-1\leq m \leq n+1$,  in particular, the case of $m=n+1$ has no internal cells. 
We solve this part of the recurrence relation  
 by setting initial values,
\begin{eqnarray}
 &&S_{-n,-i}=\beta_{n,n-1-2i} \ (i=0,1,\cdots,n-1), \\
 &&S_{-n+i,i}=\beta_{n,n-2i} \ (i=1,\cdots,n-1),
\end{eqnarray}
where $\beta_{ni}$ are any $2n-1$ complex values. It follows that  
$S_{n,n}=S_{-n,-n}=0$, $S_{n-1-i,n}=\beta_{n,n-1-2i}\ (i=0,1,\cdots,n-1)$. 
Thus, on $R_n$,
the solution can be,  {\it at least,} expressed as
\begin{eqnarray}
 && S_{-n,-i}=S_{n-1-i,n}=\beta_{n,n-1-2i},\qquad (i=0,1,\cdots,n-1),\label{S_R1}\\
 && S_{-n+i,i}=\beta_{n,n-2i}, \qquad\qquad\qquad\qquad\,  (i=1,\cdots,n-1),\label{S_R2}\\
 && S_{n,n}=S_{-n,-n}=0. \label{S_R3} 
\end{eqnarray} 
This  does  not exclude the possibility that some of $\beta_{ni}$ are equal to zero or linearly dependent of each other, because $A_{mi}=0\ (m\leq n)$ could be further constraints on them. 

\begin{figure}[t!]
\begin{center}
 {\includegraphics[width=0.4\columnwidth,clip]{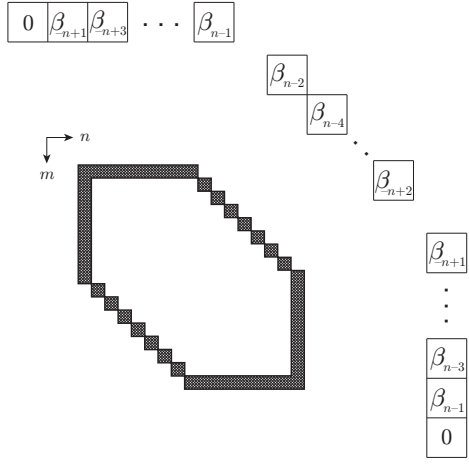}}\label{fig_R_N}\\
\caption{Boundary structure of $M_n$. The blocks having $\beta_i\equiv \beta_{ni}$ denote the terms  in top, tip-right and right of the $n$-th boundary $R_n$ (shaded region).
\label{fig_R_N_all}}
\end{center}
\end{figure}

Figure \ref{fig_R_N_all} shows this boundary structure (\ref{S_R1})--(\ref{S_R3}).
The boundary consists of the six domains (shaded region or cells): top, top-right, right, bottom, bottom-left and left in the clockwise direction. 
From the symmetry of $S$, the top (top-right/right) coincides with the left 
(bottom-left/bottom).  The top (right) corresponds to the first (second) terms in  (\ref{S_R1}) and (\ref{S_R3}),  while the top-right is given by  (\ref{S_R2}).  
Note that $\beta_{ni}$ in the top and those in the right are assigned in the opposite  order.

% fig. of R_N 
% fig. of L_{Nk}
\begin{figure}[t!]
\begin{center}
  {\includegraphics[width=0.4\columnwidth,clip]{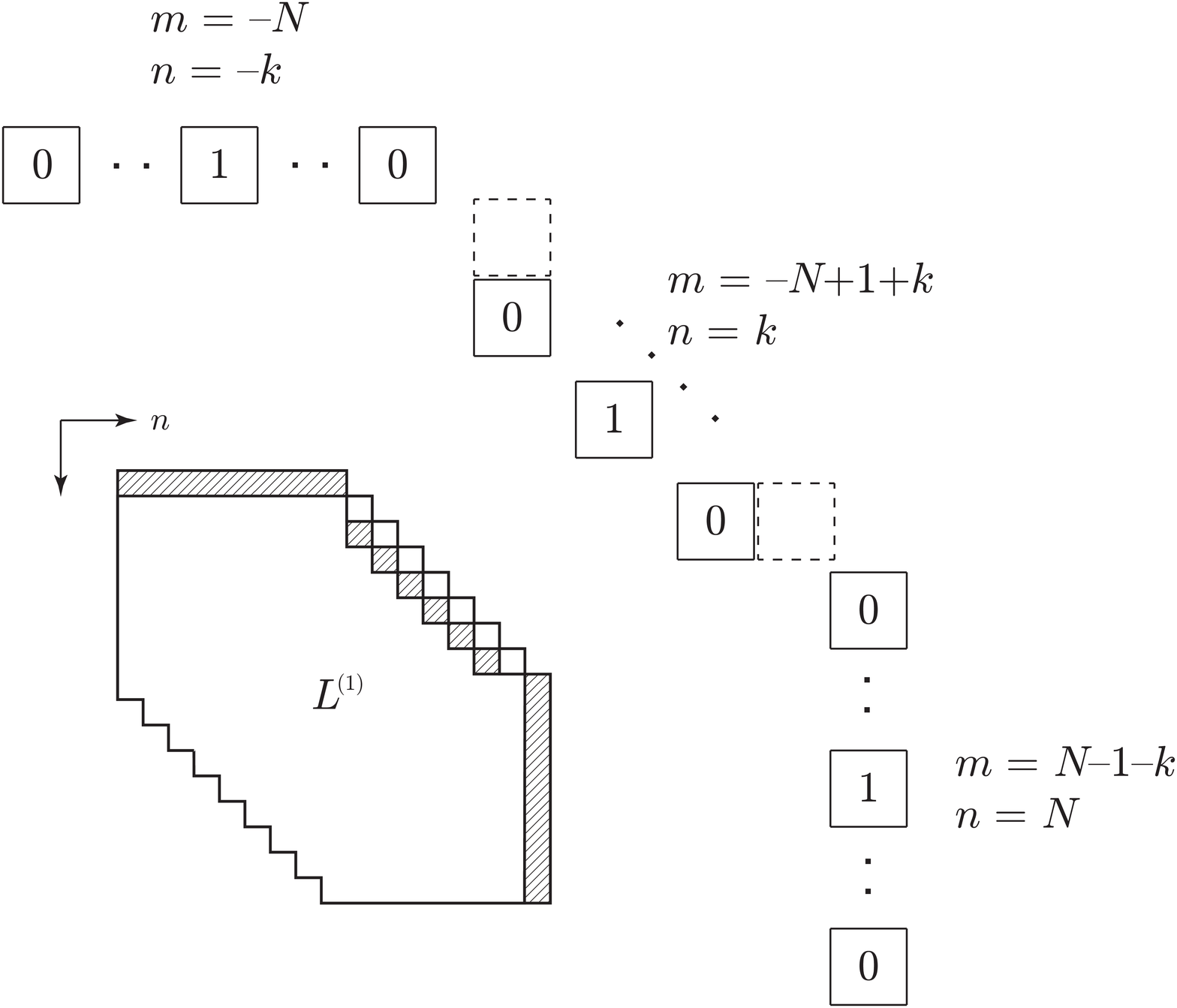}}
\hspace{1.5cm}
  {\includegraphics[width=0.4\columnwidth,clip]{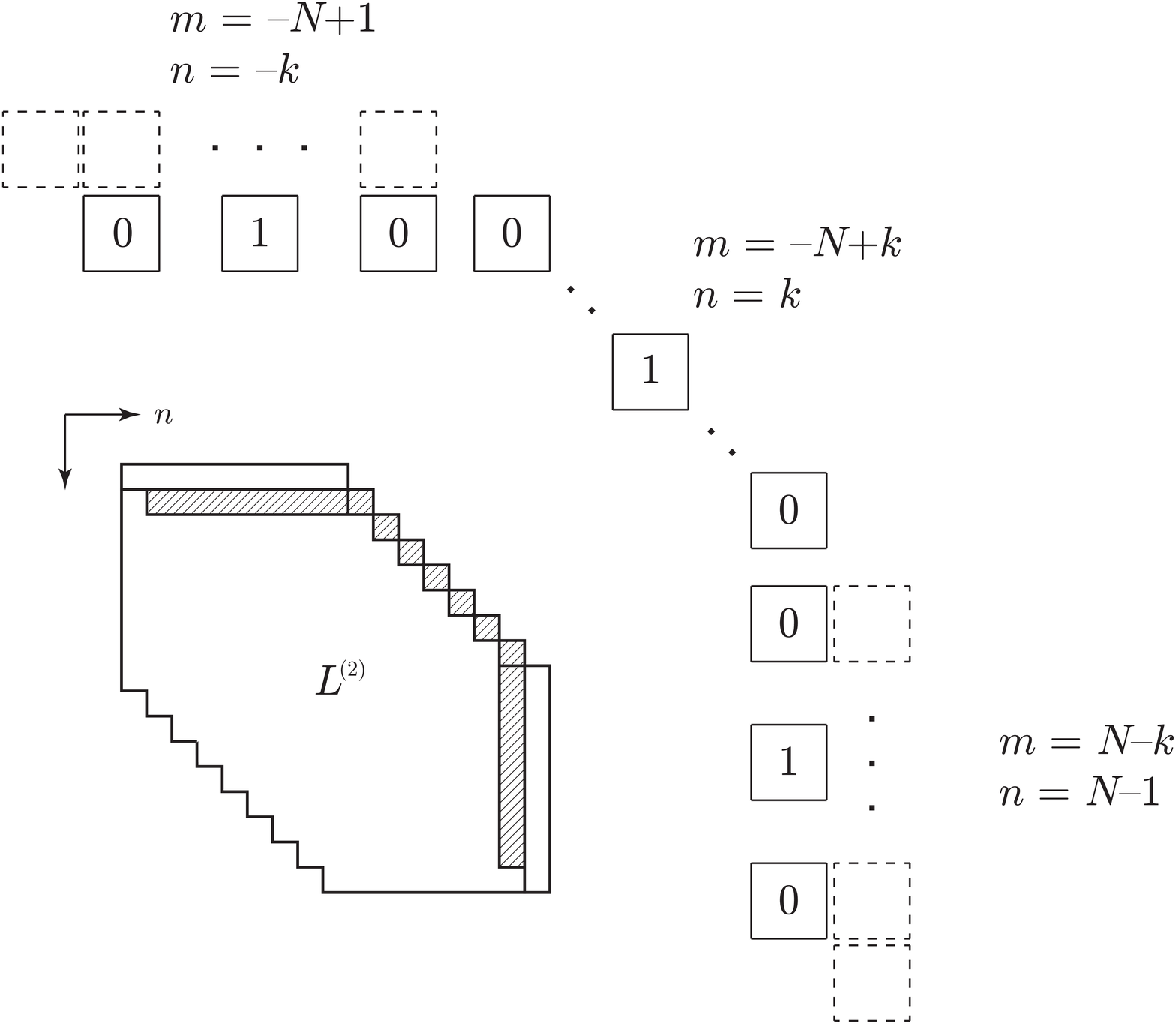}}
\caption{Table representations of $L_{ni}^{(1)}$ (left) and $\hL_{ni}^{(2)}$ (right)\label{fig_LNk}}
\end{center}
\end{figure}

By using the $n$-th degree minimal solutions (\ref{sol_min}), we can define the following $n$-th degree solution,  
\begin{eqnarray}
 \hW_n =\sum_{i=-n+1}^{n-1} \beta_{ni} \hL_{ni},
 \label{tM}
\end{eqnarray}
where $\beta_{ni}$ are given in (\ref{S_R1}) and (\ref{S_R2}).
From the table representation of the minimal solution (figure \ref{fig_LNk}), 
it is found that $\hW_{n}$ is equal to $\hM_n$ on $R_n$,  (\ref{S_R1})-(\ref{S_R3}).
Both $\hW_{n}$ and $\hM_n$ have degree $n$, but the difference
\begin{eqnarray} 
 \hM^{\prime} =\hM_{n} - \hW_n
\end{eqnarray}  
is a $n-1$-th degree solution because it vanishes on $R_n$. 
%, 
%
The lemma for $n-1$ tells us
that the $n-1$-th degree solution $\hM^{\prime}$
%,
can be written as
\begin{eqnarray}
   \hM^{\prime} =  
  \sum_{m=1}^{n-1}
  \sum_{i=-m+1}^{m-1} \beta_{mi} \hL_{mi},
\end{eqnarray}  
where $\beta_{mi}$ are the complex constants.
Finally, choosing $\beta_{mi}$ as $\alpha_{mi}$, 
$\hM_{n}$ can be written as
\begin{eqnarray}
   \hM_{n} =  
  \sum_{m=1}^{n}
  \sum_{i=-m+1}^{m-1} \alpha_{mi} \hL_{mi}.
\end{eqnarray}  
Since both the base and the inductive steps have been done,  the lemma B  holds for any finite degree $n$. \qquad $\Box$
% by mathematical induction.
\vspace{1cm}

%%%%%%%%%%%%%%%%%%%%%%%%%%%%%%%%%%%%%%%%%%%%%%%%%%%%%%%%%%%%%%%%%%%%
%                                                                                                   %
%                                                                                                   %
%                                                                                                   %
%                       Appendix. C                                                            % 
%                                                                                                   %
%                                                                                                   % 
%                                                                                                   %
%%%%%%%%%%%%%%%%%%%%%%%%%%%%%%%%%%%%%%%%%%%%%%%%%%%%%%%%%%%%%%%%%%%%

\section{Extension to general difference operators}
In this appendix, we prove that the number of fundamental solutions of the 2-body CLR is two for any difference operator.

Let us consider the general difference operators that satisfy 
\begin{eqnarray}
 \sum_{n}\D_{mn} =0
 \label{D0}
\end{eqnarray}
 in the coordinate space.
 We assume that $\D_{mn}$ are translational invariant operators which satisfy the locality condition (\ref{local_D}) 
 and become $\frac{d}{dt}$ in the continuum limit. From (\ref{difference-operator_def}), the condition (\ref{D0}) means that they vanish when acting on any constant.
Of course, the symmetric difference operators (\ref{difference-operator_sym}) satisfy these conditions.

In {\it w}-representation, the corresponding $\hD(w)$ is a holomorphic function in ${\cal D}_{K}$ which satisfies 
\begin{eqnarray}
 &&\hD(1) = 0,\\
 &&\left.\frac{\partial\hD(w)}{\partial w}\right|_{w=1} = 1.
\end{eqnarray}
There exists a holomorphic function $P_{\hD}(w)$ % that is non-zero at $w=1$
such that 
\begin{eqnarray}
 \hD(w) = (w-1)P_{\hD}(w).
\end{eqnarray}
where $P_{\hD}(1)=1$.
Let $\hM$ be a solution of the CLR for $\hD$. For $\hD^{\prime}$, we can find a solution,  
\begin{eqnarray}
%\hD^{\prime}\left(w\right) &\equiv& \hD\left(w\right)\frac{P_{\hD^{\prime}}\left(w\right)}{P_{\hD}\left(w\right)},
% \label{new_set_hD}\\
 \hM^{\prime}(w,z) &\equiv&
   \hM(w,z)
      P_{\hD^{\prime}}(w)P_{\hD^{\prime}}(z)P_{\hD}\left(\iwz\right).
       \label{new_set_hM}
\end{eqnarray}
%
%where 
%$\hD^{\prime}\left(w\right)=\left(w-1\right)P_{\hD^{\prime}}\left(w\right)$,  and 
%$\hM^{\prime}$ is also holomorphic in ${\cal C}_{K}$.
%From a short calculation, it is easy to show that the new solution pair also satisfies  the CLR. 
Note that $\hM^{\prime}$  is holomorphic and is not identically zero on ${\cal C}_{K}$.% because it is non-zero at least in an open neighborhood of $w=z=1$.

The fundamental solutions $b_i$ are  linearly independent bases of the solution space in the sense that   
$\sum_{i=1}^{m}f_i\hb_i=0\Leftrightarrow f_{i}=0$ in %any open subset of
 ${\cal C}_{K}$. 
Suppose that there are $m$ fundamental solutions $\hb_i\ (i=1,\cdots,m)$ 
for $\hD$, and $m^{\prime}$ fundamental solutions $\hb_i^{\prime}\ (i=1\cdots,m^{\prime})$ for $\hD^{\prime}$. 
Assume $m^{\prime} < m$. Then, (\ref{new_set_hM}) tells us that
\begin{eqnarray}
 \hM_i^{\prime}(w,z) = \hb_i(w,z) P_{\hD^{\prime}}(w)P_{\hD^{\prime}}(z)P_{\hD}\left(\iwz\right),
 &&\quad {\rm for\ }i=1,\cdots,m, \label{b_12m}
\end{eqnarray} 
satisfy $\sum_{i=1}^{m}f_i\hM_i^{\prime}=0$ for ${}^\exists f_{i}\neq0$.  But the identity theorem (see \cite{Gunning}) suggests us that the product of $P$ in the right hand side  of (\ref{b_12m}) is removable
and  the non-zero $f_i$ satisfy $\sum_{i=1}^{m}f_i\hb_i=0$ 
in ${\cal C}_{K}$.  Therefore the assumption $m^{\prime}<m$ must be false and $m^{\prime}\geq m$.
Conversely, the same procedure for $b_i^{\prime}$ tells us that $m^{\prime}\leq m$. Namely, $m=m'$. 
Therefore, the number of fundamental solutions does not depend on the type of the difference operators.  

Since there exist two fundamental solutions for the naive symmetric difference operator, the number of fundamental solutions for any difference operator is two.

%
% references
%

\end{document}